\documentclass[twocolumn]{pasj00}

\SetRunningHead{Y. Terashima et al.}{X-ray Spectral Variability of NGC 4051}
\Received{}
\Accepted{}

\begin{document}

\title{X-Ray Spectral Variability of the Seyfert Galaxy 
NGC 4051 Observed with {\em Suzaku}
}

\author{Yuichi \textsc{Terashima}\altaffilmark{1}, 
Luigi C. \textsc{Gallo}\altaffilmark{2},
Hirohiko \textsc{Inoue}\altaffilmark{3,4},
Alex G. \textsc{Markowitz}\altaffilmark{5},
James N. \textsc{Reeves}\altaffilmark{6},
Naohisa \textsc{Anabuki}\altaffilmark{7},
Andrew C. \textsc{Fabian}\altaffilmark{8},
Richard E. \textsc{Griffiths}\altaffilmark{9},
Kiyoshi \textsc{Hayashida}\altaffilmark{7},
Takeshi \textsc{Itoh}\altaffilmark{10},
Norihide \textsc{Kokubun}\altaffilmark{3},
Aya \textsc{Kubota}\altaffilmark{11},
Giovanni \textsc{Miniutti}\altaffilmark{12},
Tadayuki \textsc{Takahashi}\altaffilmark{3,10},
Makoto \textsc{Yamauchi}\altaffilmark{13},
and
Daisuke \textsc{Yonetoku}\altaffilmark{14}
}
\altaffiltext{1}{Department of Physics, Ehime University, 2-5 Bunkyo-cho, Matsuyama, Ehime, 790-8577}
\altaffiltext{2}{Department of Astronomy and Physics, Saint Mary's University, 
Halifax, NS B3H 3C3, Canada}
\altaffiltext{3}{
Institute of Space and Astronautical Science, Japan Aerospace
Exploration Agency, 3-1-1 Yoshinodai, Sagamihara, Kanagawa 229-8510
}
\altaffiltext{4}{
Department of Physics, Tokyo Institute of Technology, 2-12-1 Ohokayama, 
Meguro, Tokyo 152-8551, Japan
}
\altaffiltext{5}{
Center for Astrophysics and Space Sciences, University of
California, San Diego, M.C.0424, La Jolla, CA 92093-0424, USA
}
\altaffiltext{6}{
Astrophysics Group, School of Physical and Geographical Sciences,
Keele University, Keele, Staffordshire ST5 8EH, UK
}
\altaffiltext{7}{
Department of Earth and Space Science, Osaka University, Toyonaka,
Osaka 560-0043, Japan
}
\altaffiltext{8}{
Institute of Astronomy, University of Cambridge, Madingley Road,
Cambridge CB3 0HA, UK
}
\altaffiltext{9}{
Department of Physics, Carnegie Mellon University, 5000 Forbes Avenue,
Pittsburgh, PA 15213, USA
}
\altaffiltext{10}{
Department of Physics, The University of Tokyo, 7-3-1 Hongo, Bunkyo, 
Tokyo 113-0033}
\altaffiltext{11}{
Department of Electronic Information Systems, Shibaura Institute of
Technology, 307 Fukasaku, Minuma-ku, Saitama-shi, Saitama 337-8570
}
\altaffiltext{12}{
Laboratoire APC, UMR 7164, 10 rue A. Domon et L. Duquet, 75205 Paris,
France
}
\altaffiltext{13}{
Department of Applied Physics, University of Miyazaki, 1-1
Gakuen-Kibanadai-Nishi, Miyazaki, Miyazaki 889-2192
}
\altaffiltext{14}{
Department of Physics, Kanazawa University, Kakuma-machi, Kanazawa
920-1192, Japan
}


\maketitle

\begin{abstract}

We report results from a {\it Suzaku} observation of the narrow-line
Seyfert 1 NGC 4051. Broad-band X-ray light curves and spectra in the
0.4--40 keV band are examined.  During our observation, large
amplitude rapid variability is seen and the averaged 2--10 keV flux is
$8.1\times10^{-12}$ erg s$^{-1}$ cm$^{-2}$, which is several times
lower than the historical average. The X-ray spectrum hardens when the
source flux becomes lower, confirming the trend of spectral
variability known for many Seyfert 1 galaxies. The broad-band averaged
spectrum and spectra in high and low flux intervals are analyzed.  The
spectra are first fitted with a model consisting of a power-law
component, a reflection continuum originating in cold matter, a
blackbody component, two zones of ionized absorber, and several
Gaussian emission lines. The amount of reflection is rather large ($R
\sim 7$, where $R=1$ corresponds to reflection by an infinite slab),
while the equivalent width of the Fe-K line at 6.4 keV is modest (140
eV) for the averaged spectrum. We then model the overall spectra by
introducing partial covering for the power-law component and
reflection continuum independently.  The column density for the former
is $1\times10^{23}$ cm$^{-2}$, while it is fixed at $1\times10^{24}$
cm$^{-2}$ for the latter.  By comparing the spectra in different flux
states, we identify the causes of spectral variability.  At high
energies ($>$3.5 keV) the primary cause of the spectral variability is
the change of the normalization of the power-law component with a
constant photon index ($\Gamma = 2.04$) overlaid on a nearly constant
hard component.  The constant hard component is interpreted as
partially covered reflection.  At lower energies, variations in the
covering fraction of the power-law continuum is an additional cause of
spectral variability.

\end{abstract}

\section{Introduction}

NGC 4051 is an X-ray bright, nearby ($z=0.00234$) narrow-line Seyfert
1 galaxy. Its X-ray flux and spectral shape, which have been studied
with major X-ray satellites, are highly variable both on short ($<$
day) and long (years) timescales.  Such spectral variability can be
used to extract the various emission components and to understand the
physical origin of the variability.  Rapid flux and spectral variability
within a few hundred seconds have been known since observations with {\it
EXOSAT} (Lawrence et al.\ 1985) and then {\it Ginga} (Matsuoka et
al.\ 1990, Kunieda et al.\ 1992).  Extensive long-term monitoring by
{\it RXTE} showed that this object displays large amplitude variability
and occasionally enters periods of very low flux (Uttley et al.\ 1999;
Markowitz \& Edelson 2004).


The trend of the spectral variability above a few keV is that the
spectrum flattens when the flux becomes lower. In a high flux state,
its spectrum in the 0.5--10 keV band is typical of that seen for Seyfert 1s,
consisting of a power-law 
component, an Fe-K line, a reflection continuum originating
in cold matter, a soft excess, and ionized absorbers (e.g., Pounds et
al.\ 2004 for an {\it XMM-Newton} observation in 2001). When the source became
fainter, the hard band spectrum above $\sim$ 2 keV significantly
flattened and a component with a convex shape was seen (Uttley et
al.\ 2004, Pounds et al.\ 2004, Ponti et al.\ 2006, Gallo 2006
for an {\it XMM-Newton} observation
in 2002 Nov.; Uttley et al.\ 2003 for the {\it Chandra} observation in 2001
Feb.).  This component was interpreted as a relativistically broadened
Fe-K line (Uttley et al.\ 2003), a combination of reflection from an
ionized disk and a Fe-K line, both blurred by relativistic effects
(Ponti et al.\ 2006), or a primary continuum partially covered by a
weakly ionized absorber ($\xi \approx 25$ erg~cm~s$^{-1}$, 
Pounds et al.\ 2004).

NGC 4051 occasionally enters a very low flux state. Such an event was
observed by {\it RXTE} and {\it BeppoSAX} in 1998, with an X-ray flux of
$1.3\times10^{-12}$ erg s$^{-1}$cm$^{-2}$ in the 2--10 keV band. The observed
X-ray spectrum was very flat ($\Gamma \approx 0.8$) and accompanied by a
strong Fe-K line with an equivalent width ($EW$) $\approx 600$ eV, consistent
with a reflection-dominated spectrum. This spectrum is
consistent with the idea that the nuclear activity had switched
off and that the observed X-rays were
reflected from matter located far from the nucleus and possibly
identified with the molecular torus (Guainazzi et al.\ 1998, Uttley et
al.\ 1999).

%
%
Several different interpretations for the observed spectral variability
have been proposed. A first possibility is that the spectral
variability is mainly caused by an intrinsic change of the power law
slope.  Guainazzi et al.\ (1996) analyzed spectral variability using
{\it ASCA} data obtained in 1994 and found that the power law slope
changed by $\Delta \Gamma = 0.4$ if the normalization of the
reflection component is constant or inversely proportional to the
normalization of the power-law component.  Lamer et al.\ (2003) analyzed
long-term spectral variability using {\it RXTE} data and found that the
slope of the power-law component varies even after subtracting a
reflection component, which is assumed to be constant.  Taylor et
al.\ (2003) performed flux-flux analysis for {\it XMM-Newton} data using the
two energy bands 2--5 keV and 7--15 keV, and concluded that the
spectral variability is described by a pivoting power-law component
(i.e., change of the power law slope) superimposed on
a constant hard component. Uttley
et al.\ (2004) analyzed flux-flux plots in the 0.1--0.5 keV and 2--10
keV bands for high and low flux states observed with {\it XMM-Newton} in 2002
Nov.\  and reached a similar conclusion. They explained the spectral
variability by a power-law component with a varying photon index combined with
constant thermal and reflection components, confirming the
interpretation of {\it Chandra} results by Uttley et al.\ (2003).

%
%
A second scenario is a two-component model consisting of a soft
variable component and a hard constant component. Ponti et al.\ (2006)
explained the spectral variability observed with {\it XMM-Newton} in 2001 and
2002 with this two-component model, in which
a power law with a fixed
photon index ($\Gamma = 2.2$) was used
as the variable component. A combination
of ionized reflection blurred by a Laor (1991) kernel and neutral
reflection from distant matter was introduced as the constant component. The
former is interpreted as constant in the context of the light bending
effect around the central Kerr black hole (Miniutti \& Fabian 2004).

%
%

Pounds et al.\ (2004) analyzed the same {\it XMM-Newton} data and interpreted the
variability using the two-component model combined with the presence of
an absorbed component.  They assumed that the strength of the
reflection component and the photon index of the power-law component
were same between the two observations and interpreted the very
hard spectrum observed in 2002 by a combination of a
significant contribution of a 'constant' cold reflection component and
an increased column density of absorption  with
a low ionization state.


Spectral variability has been thus interpreted by various combinations
of a pivoting power-law component, 
a constant flat component, and absorption, and
the underlying physics is yet to be understood. A powerful way to
decompose these various components and to extract the variable component(s) is
time-resolved broad-band spectroscopy.  NGC 4051 is bright at energies
above 10 keV and detected by {\it Ginga} (up to 20 keV; e.g., Matsuoka et
al.\ 1990, Awaki et al.\ 1991, Kunieda et al.\ 1992), {\it BeppoSAX} (up to 50
keV in a low flux state, Guainazzi et al.\ 1998), {\it INTEGRAL} (over
17--60 keV, Sazonov et al.\ 2007, Krivonos et al.\ 2007, Beckmann et
al.\ 2006), and {\it Swift} (in the 14--195 keV band, Tueller et al.\ 2008).
It is therefore one of ideal targets to study the nature of
broad-band spectral variability with {\it Suzaku}, which achieves
broad-band coverage up to 40 keV with a good signal-to-noise ratio for
sources with a moderate flux. Indeed, the variable emission component in
Seyferts has been successfully measured in {\it Suzaku} observations
to be a power law with a constant photon index 
(MCG--6-30-15, Miniutti et al.\ 2007; MCG--5-23-16,
Reeves et al.\ 2007; NGC 4388, Shirai et al.\ 2008; NGC 4945, Itoh et
al.\ 2008).

  In this paper, we present the results of analysis of spectral
variability of NGC 4051 using {\it Suzaku} to decompose the
variable and constant emission components.  This paper is organized as
follows:  Section 2 describes the observation and data
reduction. Section 3 and 4 describe light curves and spectra, and
the results are discussed in Section 5. Section 6 summarizes our
findings. 


\section{Observation and Data Reduction}

  NGC 4051 was observed with {\it Suzaku} on 2005 Nov.\ 10--13 as a part
of the Science Working Group observations.  {\it Suzaku} (Mitsuda et
al.\ 2007) carries four X-ray telescopes (XRTs, Serlemitsos et
al.\ 2007), with four X-ray CCD cameras (X-ray Imaging Spectrometer XIS,
Koyama et al.\ 2007) at the focal plane of the XRTs, and a non-imaging
Hard X-ray Detector (HXD, Takahashi et al.\ 2007).  Three of the four
XISs (XIS0, 2, 3) are front-illuminated (FI) CCDs and the other one
(XIS1) is a back-illuminated CCD, and has larger quantum efficiency at
lower energies compared to the FI CCDs.  The HXD consists of Si PIN
diode (PIN) and GSO scintillators surrounded by BGO active
shields. The XIS and HXD are co-aligned and NGC 4051 was placed at the
HXD nominal pointing position in this observation.


 The four XISs were operated in 5$\times$5 or 3$\times$3 editing modes
combined with normal clocking mode.  We used cleaned events from the
version v1.2 pipeline processing data.  Time intervals in which
telemetry saturation occurred were excluded by applying the good time
intervals provided by the XIS team.  Since NGC 4051 is highly variable
and a main purpose of our analysis to understand the spectral
variability in this object, we selected only time intervals where data
from the XIS and HXD/PIN were taken simultaneously.  The effective
exposure after the data screening was 81.6 ksec for each XIS.  The
mean count rates are 0.46 counts s$^{-1}$ per each FI sensor and 0.69
counts s$^{-1}$ per BI sensor in the 0.4--10 keV band. Response
matrices and ancillary response files were made for each XIS using {\tt
XISRMFGEN} and {\tt XISSIMARFGEN}, respectively.  Light curves and
spectra were extracted from a circular region with a radius of
2$^\prime$.9 arcmin.  Background spectra were accumulated from an
annular region centered on the target and a region around a nearby
source was excluded.



  The cleaned event file from the version 1.2 pipeline processing was
used for analysis of the HXD data. In this paper we analyze only the
data from the PIN because no significant signals were detected from
the GSO.  Deadtime correction was done for PIN spectra and light
curves.  Only the data taken simultaneously with the XIS were used.
The effective exposure after data screening was 81.6 ksec (before
deadtime correction) or 76.6 ksec (after deadtime correction). The
response matrix appropriate for the date of this observation and
version 1.2 processing (ae\_hxd\_pinhxnom\_20060814.rsp) was used in
spectral fits.


\begin{figure}[t]
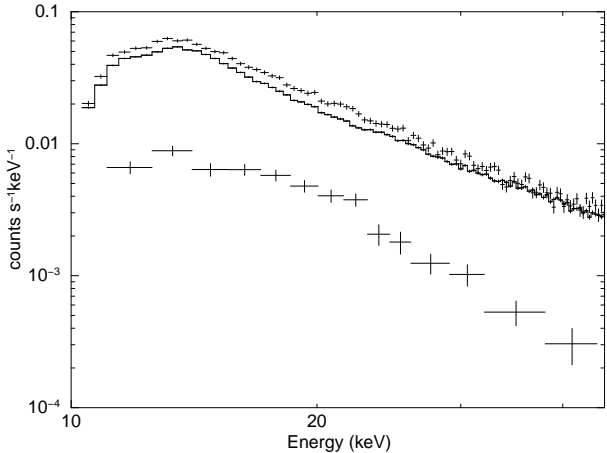

\begin{center}
\FigureFile(80mm,50mm){figure1.ps}
\caption{
HXD/PIN spectra of NGC 4051. The total observed spectrum is denoted
by the upper data points. The non-X-ray
background + Cosmic X-ray background are denoted by the middle 
data points, using a 
solid histogram. The lower data points denote the 
background-subtracted spectrum.
}
\end{center}
\end{figure}

  The non-X-ray background (NXB) appropriate for the version 1.2 products
provided by the HXD team was subtracted from spectra and light
curves, after the same time intervals
as the actual data was selected.
  We assumed the shape of the spectrum of the cosmic X-ray background
(CXB) presented in Boldt (1987) and Gruber et al.\ (1999) and simulated
an expected spectrum of the CXB using the response file for a flat
field with an  area of $2^\circ \times 2^\circ$
by using XSPEC. The form of the CXB is expressed as 
$9.0\times10^{-9} (E/3~{\rm keV})^{-0.29} \exp(-E/40~{\rm keV})$
erg cm$^{-2}$ 
s$^{-1}$ 
str$^{-1}$ 
keV$^{-1}$, or
$9.41\times10^{-3} (E/1~{\rm keV})^{-1.29} \exp(-E/40~{\rm keV})$
photons cm$^{-2}$ 
s$^{-1}$ 
keV$^{-1}$
$(2^\circ \times 2^\circ)^{-1}$.
The spectra of the NXB and CXB were combined, and then subtracted from
the on-source spectrum.  The spectra of the NXB+CXB and data before
and after background subtraction are shown in Fig.\ 1.  The count rates
of the signal are more than 10\% of those of the NXB up to $\sim40$
keV.  The uncertainties of flux measurements with HXD/PIN are
dominated by the systematic error of the NXB estimation. For example,
the 1$\sigma$ systematic error for an observation with a duration of one
day is $\approx 2$\% if statistical errors are subtracted in
quadrature (Fig.\ 3 in Mizuno et al.\ 2007). Detection is therefore
significant at more than $5\sigma$ level in the 15--40 keV band.
The accuracy of the background subtraction can be tested by comparing the data
taken during the field of view is occulted by the earth with the
NXB. Such time intervals, however, did not exist in this particular
observation. We therefore assumed the systematic error of the NXB
estimation given in Mizuno et al.\ (2007) without any further tuning.

\section{Variability}

\subsection{Light Curves}


\begin{figure}[t]
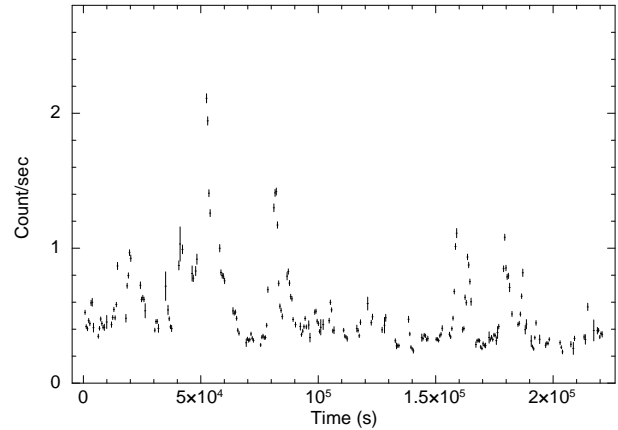

\begin{center}
\FigureFile(80mm,50mm){figure2.ps}
\caption{
The light curve in the 0.5--10 keV band obtained with the XIS, using
a bin size of 512 sec. The data from the four XIS sensors are combined.
}
\end{center}
\end{figure}

\begin{figure}[h]
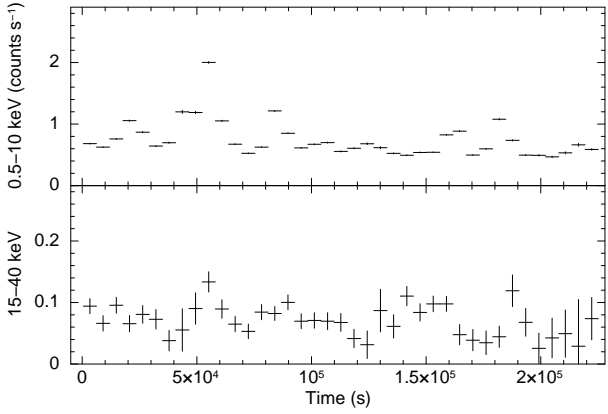

\begin{center}
\FigureFile(80mm,70mm){figure3.ps}
\caption{
(Upper) The light curve in the 0.5--10 keV band obtained with the XIS.
The bin size is 5760 sec.  The data from the four XIS sensors are
combined.
(Lower) The light curve in the 15--40 keV band obtained with the HXD PIN.
The bin size is 5760 sec. The deadtime has been corrected. The NXB has been 
subtracted. 
The 1$\sigma$ systematic error on the PIN count rates is 
$\sim 0.017$ c~s$^{-1}$.
}
\end{center}
\end{figure}

  The light curve in the 0.5--10 keV band obtained with the XIS is
shown in Fig.\ 2.  The bin size is 512 sec, and the data from the four
XIS sensors are combined.  The origin of time is 2005 Nov.\ 10,
20:12:02 (UT). During our observation, NGC 4051 showed several
flare-like time variations within a few
thousand seconds. The amplitudes of these flares are about a factor of
2--5.

  The light curve in the 15--40 keV band obtained with the HXD PIN is
compared with that in the 0.5--10 keV in Fig.\ 3. The XIS light curve
in the 0.5--10 keV band and PIN light curve in the 15--40 keV band are
shown in the top and bottom panels, respectively.  The bin size is
5760 sec and the origin of time is 2005 Nov.\ 10, 20:55:43 (UT).
Deadtime correction was done for the PIN light curves.  The hard X-ray
flux from the Cosmic X-ray background ($\sim$ 0.013 ct s$^{-1}$ in the
15--40 keV band) was not subtracted, while the NXB was subtracted in
the lower panel of Fig.\ 2. Only statistical errors are included as the
error bars. The systematic error is about 6.0\% of the NXB flux level
(0.28 ct s$^{-1} \times 0.06 \approx$ 0.017 c s$^{-1}$) at 1$\sigma$
for a time bin of $\sim$ 6 ksec (Mizuno et al.\ 2007). It is clear
from the figure that the variability amplitude is much lower in the
harder band and that the flux in the hard band is only mildly
variable.


\begin{figure}[h]
\begin{center}
\FigureFile(80mm,50mm){figure4a.ps}
\FigureFile(80mm,50mm){figure4b.ps}
\end{center}
\end{figure}

\begin{figure}[h]
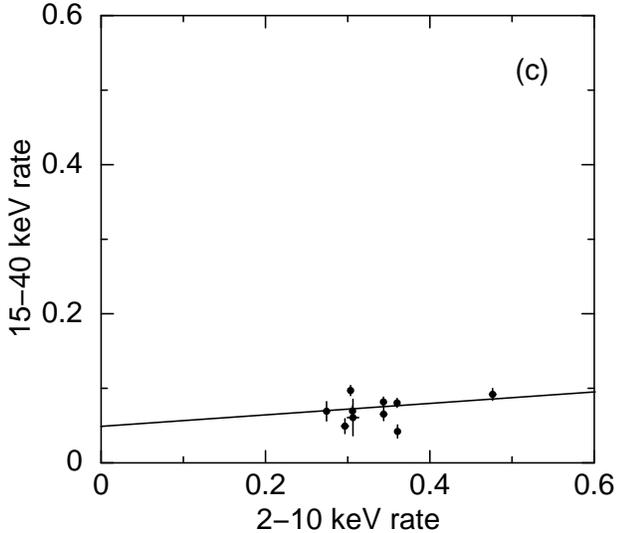

\begin{center}
\FigureFile(80mm,50mm){figure4c.ps}
\caption{
Flux-flux plots. The solid and dashed lines represent best-fit
 linear and power-law functions, respectively.
(a) 4--10 keV and 0.5--2 keV bands. The time bin size is 5760 sec. 
(b) 4--10 keV and 2--4 keV bands.  The time bin size is 5760 sec.
(c) 15--40 keV and 2--10 keV bands.  The time bin size is 23040 sec.
}
\end{center}
\end{figure}

\subsection{Flux-Flux Correlations}

  In order to characterize the spectral variability, we made flux-flux
plots for several energy bands. In Fig.\ 4(a) and (b), the count rates
in the 4--10 keV band are plotted against the count rates in the
0.5--2 keV and 2--4 keV bands, respectively, while the 15--40 keV and
2--10 keV bands are compared in Fig.\ 4(c).  The data from the four XIS
sensors were combined. A bin size of 5760 sec was used for Fig.\ 4(a) and
(b), while the bin size for Fig.\ 4(c) was 23040 sec.  These figures show
that the amplitude of variability becomes smaller as the energy band
becomes harder.
Such trends are also demonstrated by root-mean-square spectra, 
shown in the next subsection.

 We fit the flux-flux plots with a linear function and found that
there is an offset in the hard (4--10 keV) band. The offsets for Fig.\
4(a) and (b) are 0.056 ct~s$^{-1}$ and 0.046 ct~s$^{-1}$,
respectively. The best-fit lines are shown in Fig.\ 4.  Under the
assumption that a linear relationship accurately characterizes the
intrinsic flux-flux relation, the hard offset indicates that there is
a constant component with a hard spectrum and that variability is
mainly caused by a change of normalization of a spectral component of
nearly constant spectral shape. Such a two-component behavior has been
suggested based on {\it XMM-Newton} observations except for time
intervals with a very low flux (Ponti et al.\ 2006).

The data points in Fig.\ 4(a) may show a slight convex curvature
compared to the best-fit linear solution. We therefore compared the
data to a model consisting of a power law and obtained a better
description ($\chi^2$ = 495 (power law) versus 554 (linear) for 37
degrees of freedom).  The exponent of the power law is 0.54, which is
near the value 0.642 obtained by Taylor et al.\ (2003) for the two
energy bands 2--5 keV and 7--15 keV using {\it RXTE}.  The best-fit
models are also shown in Fig.\ 4(a).  Under the assumption that the
intrinsic flux-flux relation is characterized by a power law, this
result would suggest that the main cause of spectral variability
depends on the energy band and that the change of the intrinsic
spectral shape is play a role in addition to the two-component
model. Taylor et al.\ (2003) and Uttley et al.\ (2004) applied a power
law type model to their flux-flux plots, and spectral pivoting was
implied to be the main cause of spectral variability.
The comparison between the data points and the model
functions shown in Fig.\ 4 (a) and (b), along with the very large $\chi^2$
values, suggest that neither linear nor power-law models offer a
complete description of the observed flux-flux relation. Indeed,
detailed spectral analysis presented below demonstrates the importance of
modeling the variability of the absorbing material,
particularly at energies below $\sim$ 4 keV.

A flux-flux plot for harder energies is shown in Fig.\ 4(c) and
indicates that the amplitude of variability is very small in the hard
band 15--40 keV.  Note that the 1$\sigma$ systematic error in
estimating the HXD background is about 4\% of the NXB, which
corresponds to 0.01 ct s$^{-1}$, for a duration of $\sim$20 ksec
(Mizuno et al.\ 2007), and that the true amplitude is smaller than the
scatter in the vertical direction in Fig.\ 4(c). This flux-flux plot
is consistent with the two-component picture, in which the harder
X-ray band is dominated by a nearly constant component with a hard
spectrum.

\begin{figure}[h]
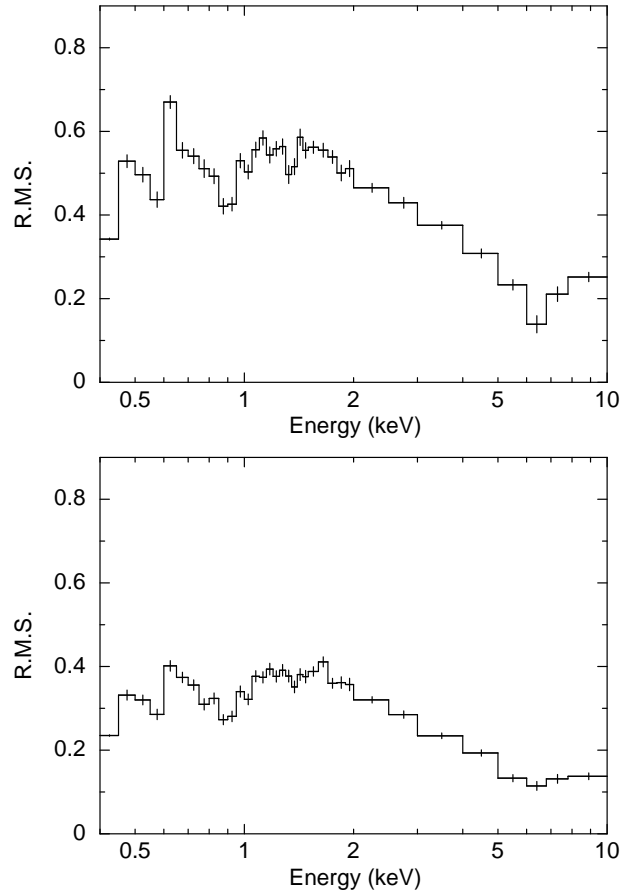

\begin{center}
\FigureFile(80mm,50mm){figure5a.ps}
\FigureFile(80mm,50mm){figure5b.ps}
\caption{The root mean square spectrum.
(a) The time bin size is 1024 sec.
(b) The time bin size is 20480 sec.
}
\end{center}
\end{figure}

\subsection{RMS spectra}


Fractional variability amplitudes in excess of the noise were calculated
for light curves in narrow energy bands and plotted against
energy. Such an RMS spectrum is a useful way to quantify spectral
variability (e.g., Vaughan et al.\ 2003, Markowitz et al.\ 2003). We
used the definition and error of the fractional root mean square
variability amplitude ($F_{\rm var}$) shown in Vaughan et al.\ (2003).
RMS spectra were calculated for two time bin sizes, 1024 and 20480
sec, and are shown in Fig.\ 5.  The overall shape is such that the RMS value is
peaking at around $\sim 0.6-1$ keV and then gradually decreases toward
higher energies.  Such a shape peaking near 1--2 keV is known for some
Seyfert 1 galaxies (MCG--6-30-15, Vaughan \& Fabian 2004; Mrk 766,
Ar\'{e}valo et al.\ 2008; 1H0707$-$495, Fabian et al. 2004, Gallo et al.
2004; IRAS~13224$-$3809, Gallo et al. 2007).  
Several sharp drops are also clearly seen
at 0.58, 0.90, 1.35, and 6.4 keV, thanks to the good energy resolution of
the XIS at low energies and the high variability-to-noise ratios in
this source.  These energies coincide with emission lines seen in the
spectra as shown below and can be identified with K-shell lines from 
O VII, Ne IX, Mg XII, and neutral or low ionization Fe, respectively.  Such
sharp features are not clear in an RMS spectrum measured from
a high flux state observation in 2001, while a 
dip at 0.9 keV has been observed in an RMS spectrum of a low
flux state in 2002 (Ponti et al.\ 2006). Other lines are less clear in
the previous RMS spectra, though localized
decreases in the RMS around 0.5 keV, 1.3 keV,
and 6.5 keV are seen.

\section{Spectral Variability}

We divided the data into high and low flux states to study spectral
variability, where a threshold of 0.65 counts s$^{-1}$ per XIS0 was
used. In order to extract variable components, we first made a
difference spectrum between the high and low flux states, and then
examined spectra of each flux level (high and low) as well as
the averaged spectrum.
The three XIS-FI spectra were co-added. Response matrices and ancillary
response files for these three XIS-FI sensors were combined using {\tt
ADDRMF} and {\tt ADDARF}, respectively.  We fitted the XIS-FI,
XIS-BI, and PIN spectra simultaneously, though only XIS-FI and PIN
data are shown in figures below for clarity, unless otherwise noted. In
fitting XIS and HXD PIN spectra, we assumed that the relative
normalizations of XIS0+2+3, XIS1, and PIN was 1:0.96:1.16 based on
version 1.2 data of the Crab nebula analyzed by the {\it Suzaku} X-ray
Telescope team.

The XIS and HXD/PIN background-subtracted spectra were fitted using
XSPEC v11.3. The spectra were binned so that each bin contains at
least 30 counts and the chi-squared minimization technique could be used to
fit the spectra. The energy band 1.8--1.9 keV of the XIS spectra was
ignored in the spectral fits to avoid residual calibration
uncertainties known in this band.  The hydrogen column density of the
Galactic absorption $N_{\rm H}$ = $1.2\times 10^{20}$ cm$^{-2}$ (Kalberla et
al.\ 2005) obtained by the tool {\tt nh} in the HEASOFT 6.3.1 package
was included in the spectral models described below. All errors quoted
are at 90\% confidence for one parameter of interest ($\Delta \chi^2$
= 2.7).

Spectra of the $^{55}$Fe calibration sources in the XIS cameras
were extracted
to quantify residual calibration uncertainties.
We fitted the spectra around the peak of the Mn-K$\alpha$ line 
with a Gaussian component and measured the peak energy and width $\sigma$.
The measured line energies were within 5 eV of the expected energy. The widths
were consistent with zero. The largest residual width is seen in XIS3
($\sigma \approx 11$ eV). We therefore estimate that the uncertainties in
the energy and width are less than $\sim$ 5 eV and $\sim$ 10 eV
for the simultaneous fits presented below.


\begin{figure}[h]
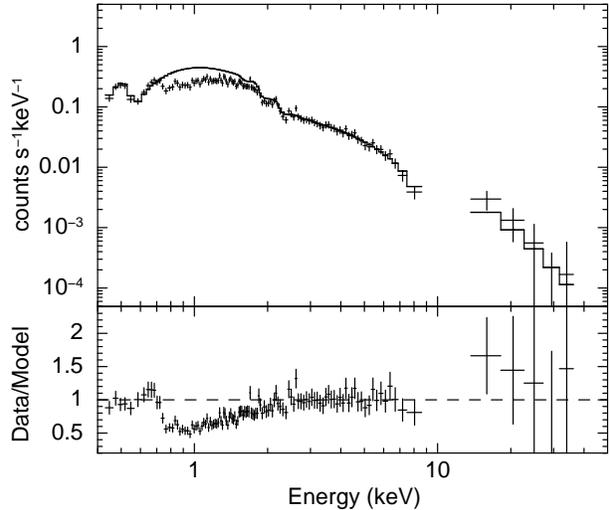

\begin{center}
\FigureFile(80mm,50mm){figure6.ps}
\caption{
The difference spectrum between the high and low flux states. XIS 1 data
are not shown for clarity, though they were used in the spectral fits.
The upper panel shows the difference spectrum (crosses) and the
best-fit power law model in the 3--40 keV band extrapolated to
energies below 3 keV (histogram). The lower panel shows 
data/model ratio.
}
\end{center}
\end{figure}

\subsection{Difference Spectrum}

The spectrum in the low flux state was subtracted from that in the
high flux state to make a difference spectrum.  Since the background
of the XIS is stable and similar background count rates are expected for
spectra in both states, we used XIS spectra without background
subtraction. Errors in each spectral bins were propagated.  On the
other hand, the background of the PIN is highly time-dependent, and
therefore we prepared background-subtracted spectra for both states,
and then a difference spectrum was made using them. Errors were
propagated in each arithmetic operation.

The difference spectrum thus obtained is shown in Fig.\ 6. We first
fit this spectrum using a simple power law model without any
modifications by absorption in the 3--40 keV band.  The spectrum is
well fitted with a power law with a photon index of $2.04\pm0.16$
($\chi^2$/dof was 18.8/47) and no systematic residuals are seen.  The
best-fit power law model extrapolated to lower energies is also shown
in Fig.\ 6. Absorption due to the Galactic column ($N_{\rm H}$ =
$1.2\times10^{20}$ cm$^{-2}$) is applied in this figure. The data in
the 0.7--2 keV band are located below the extrapolation of the power
law and indicate the presence of absorption in this energy band.


 Next we added a narrow ($\sigma = 10$ eV) Gaussian at 6.4 keV in the
source rest frame to set a limit on the variability of the Fe-K 
emission line. The
90\% confidence upper limit of the equivalent width ($EW$) 
is 160 eV. The presence of a reflection
component was also examined. We used the PEXRAV component in XSPEC 
(Magdziarz \& Zdziarski 1995) to constrain the strength of the 
reflection ($R$ or {\tt rel\_refl} in the PEXRAV component) from cold
matter. $R$ is the reflection fraction, defined as $R = \Omega / 2\pi$,
where $\Omega$ is the solid angle of a cold reflector viewed from an
X-ray source. An inclination angle of $i=45^{\circ}$ and an incident
spectrum consisting of a power law with an exponential cutoff energy of 300 keV
were assumed in the fits. The photon index and $R$, however, are
strongly coupled with each other and no meaningful constraint was
obtained. 
If we fixed the photon index at 2.04, the upper limit on $R$
became 1.7 ($\Delta \chi^2 = 2.7$) or 2.1 ($\Delta \chi^2 = 4.6$),
which are consistent with the absence of strong reflection,
although these constraints are not tight.

\begin{figure}[h]
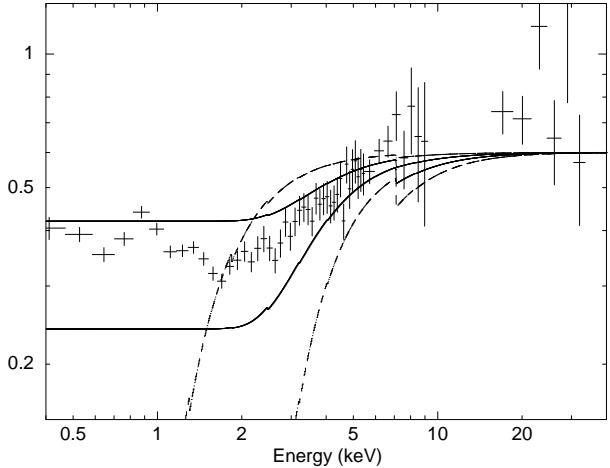

\begin{center}
\FigureFile(80mm,50mm){figure7.ps}
\caption{ Ratio of the low flux state to high flux state spectra.  The
upper and lower dashed lines are the ratios expected for absorption by
neutral matter with a column density of $N_{\rm H} = 10^{22}$
cm$^{-2}$ and $10^{23}$ cm$^{-2}$, respectively.  Upper and lower
solid lines are the ratios expected for partial covering absorption
with covering fractions of 0.3 and 0.6, respectively, where a column
density of $N_{\rm H} = 10^{23}$ cm$^{-2}$ is assumed.  In these
models, the continuum flux is assumed to be reduced by 40\% in the low
flux state.  }
\end{center}
\end{figure}

\subsection{Spectral Ratio}

The difference spectrum in the previous section suggests the presence
of variable absorption. Since absorption is a multiplicative
component, a ratio of the low flux to high flux spectra is useful to
see the effect of absorption. We therefore made such a ratio plot (low
flux/high flux states), shown in Fig.\ 7. A few absorption models are
also shown in Fig.\ 7, in which we assumed that the shape of the
underlying continuum remains the same, and the normalization of the
continuum is reduced by 40\% in the low flux state compared to the the
high flux state.  The upper and lower dashed lines in Fig.\ 7 are the
ratios expected for absorption by neutral matter with a column density
of $N_{\rm H} = 10^{22}$ cm$^{-2}$ and $10^{23}$ cm$^{-2}$,
respectively. The energy dependence of the ratio data is shallower
than these curves in that the data points lie much higher than the
models at low energies (below 2 keV). This fact indicates that a
simple neutral absorber does not explain the observed spectral
variability.  The ratios expected for partial covering absorbers with
covering fractions of 0.3 and 0.6 are also shown as upper and lower
solid lines, respectively, in Fig.\ 7, where a column density of
$N_{\rm H} = 10^{23}$ cm$^{-2}$ was assumed. Both sets of model points
are nearly flat at energies below $\sim$ 3 keV. The fact that they are
similar in spectral shape to the observed ratio spectrum at low
energies suggests the presence in NGC 4051 of less absorbed
components, which might be modeled using partial covering.  Full
details of absorption models are explored in the following
subsections.

\begin{figure}[h]
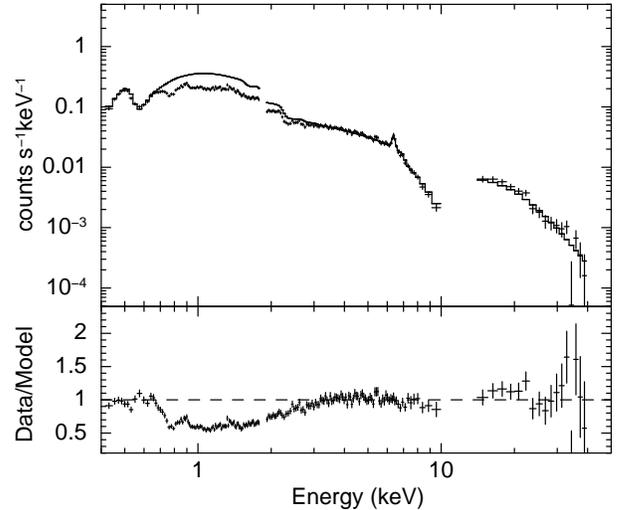

\begin{center}
\FigureFile(80mm,50mm){figure8.ps}
\caption{
The averaged spectrum fitted with the two-component model 
in the 3.5--40 keV band.
The best-fit model (solid histogram) is extrapolated to
lower energies.
The lower panel shows data/model ratio.}
\end{center}
\end{figure}

\subsection{Spectra in the 3.5--40 keV band}

\subsubsection{Empirical Two-component Model}


The result of spectral fits to the difference spectrum in the previous
subsection suggests that the variable component in the hard energy
band above 3 keV is dominated by a power-law component, and that the
total spectrum consists of a power-law component plus
a harder constant component. We
adopt a reflection continuum expected from cold matter, which is often
present in spectra of Seyfert 1s, as the hard component. The spectra
in the two flux states were fitted with a model consisting of a 
power-law component 
and a reflection continuum. The photon index of the power-law
component was fixed at 2.04, determined by the fit to the difference
spectrum.  The PEXRAV component in XSPEC was used as the shape of the
reflection continuum, with the 
parameter {\tt rel\_refl} fixed at $-1$, meaning that 
this component was used purely to model reflection and not the
flux associated with the incident power-law component.
Gaussian components were added if 
emission/absorption features were seen. This model hereafter is
referred to as Model-A1.  The shape of the continuum is well fitted
with this model for the flux states analyzed here.  The data and
best-fit model for the averaged spectrum is shown in Fig.\ 8 as an
example. The model extrapolated to lower energies is also presented in
the same figure.  The spectral parameters of the fits in the 3.5--40
keV band are summarized in Table 1.

The normalization of the power-law component varied significantly
among the different flux states, while that of the reflection continuum
remains nearly constant. Thus the two-component model consisting of a
variable power law and constant reflection continuum represents the
apparent change of the spectral slope in the hard band.  The
extrapolation of the best-fit model towards lower energies shows
significant residuals at energies below 3.5 keV suggesting the
presence of absorption, which is explored in subsequent analyses below.
An emission line feature is seen at around 6.4 keV.  This line was
modeled with a Gaussian, and the best-fit parameters are also shown in
Table 1. The best-fit Gaussian energy centroid
indicates that the line is identified
with a K-shell line from neutral or weakly-ionized Fe.  The
normalization of this component remains constant within errors between
the various flux states.  The
width of this line is relatively narrow: it is consistent with zero
width for all the spectra.



\begin{figure}[h]
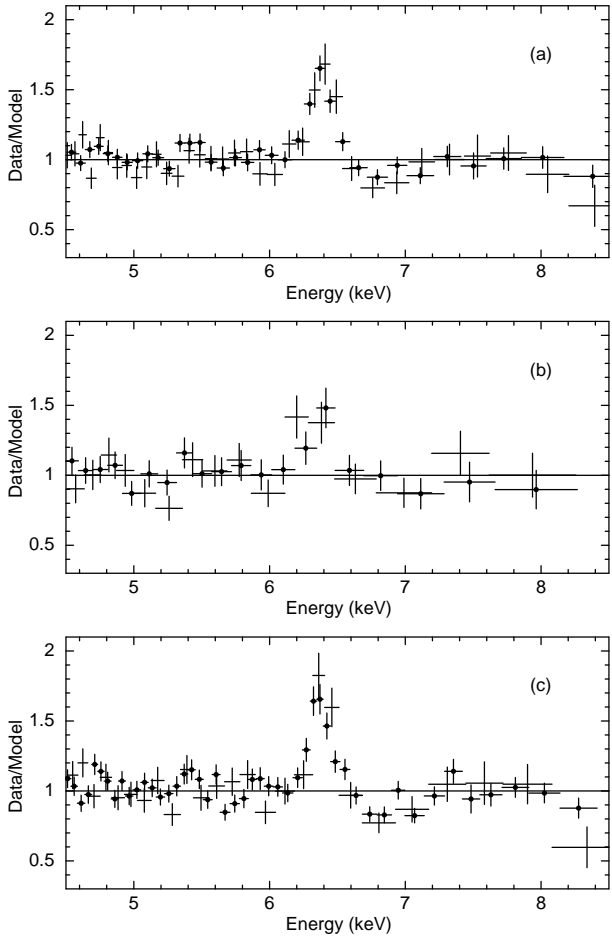

\begin{center}
\FigureFile(80mm,50mm){figure9a.ps}
\FigureFile(80mm,50mm){figure9b.ps}
\FigureFile(80mm,50mm){figure9c.ps}
\caption{
The data/continuum model ratio around the Fe-K emission line in the
averaged (a), high flux (b), and low flux (c) spectra.
XIS-FI (crosses with filled circles) and XIS-BI (crosses without
circles) data are shown. The continuum is the best-fit emprical 
two-component model fitted in 3.5--40 keV, as shown in Section 4.2.
}
\end{center}
\end{figure}

A line-like absorption feature is seen around 7 keV in the averaged
spectrum and the spectrum in the low-flux state. A Gaussian component
with a negative normalization was added to represent this feature.
The Gaussian width was fixed at $\sigma = 10$ eV. The result of this
fit is also shown in Table 1. The Gaussian energy,
$6.82^{+0.06}_{-0.10}$ keV for the averaged spectrum and $6.80\pm0.04$
keV for the spectrum in the low flux stete, is consistent with an
absorption line by He-like Fe if an outflow velocity of $\sim$ 4600 km
s$^{-1}$ is assumed. A hint of an additional absorption line is
present in the spectrum in the low-flux state.  When a narrow Gaussian
component was added, the value of $\chi^2$ was reduced by $\Delta
\chi^2$ = --10.2 for two additional parameters (centroid energy and
normalization).  The line energy, $7.10^{+0.05}_{-0.06}$ keV, is
identified with H-like Fe with an outflow velocity of $\sim$ 6000 km
s$^{-1}$ or a He-like resonance line with an outflow velocity of
$\sim$ 17000 km s$^{-1}$.  Since the line centroid energy, 7.10 keV,
coincided with the energy of the absorption edge for neutral Fe, we
examined whether this feature is truly an absorption line or an
artifact caused by improper modeling of an Fe edge absorbing the
continuum. We used an edge model instead of a Gaussian in absorption;
the edge energy was fixed at 7.1 keV in the source rest frame. The
improvement of $\chi^2$ by adding an edge was $\Delta \chi^2 = -4.1$,
and we obtained a slightly worse value of $\chi^2$, 146.9, compared to
the Gaussian case.  We also tried to change the Fe abundance in the
reflection model instead of introducing a Gaussian or an edge. The
best-fit abundance was $1.35^{+0.30}_{-0.28}$, with an improvement of
$\Delta \chi^2 = -4.8$.  Thus if there is an additional edge, the
significance of the 7.1 keV line becomes weak.

An Fe-K$\beta$ line (7.06 keV in the source rest frame, or 7.04 keV in
the observed frame) is not clearly seen in the spectra.  This is
probably due to the presence of absorption lines around the energy of
the Fe-K$\beta$ line.  The spectra in the Fe-K bandpass are shown in
Fig.\ 9, where the normalizations of the Gaussians are set to zero.

In the empirical two-component model, the reflection component
significantly contributes to the hard X-ray flux, and estimation of
the strength of the reflection is affected by the uncertainties
in the estimation of the NXB of the HXD PIN. We
examined the effect of the uncertainties in the
following way.  We increased or decreased the level of the NXB by 4\%,
which is a 90\% confidence uncertainty of the NXB estimation for a 40
ksec exposure, and refitted the averaged spectrum with the same model
described above.  The best-fit $R$ value is $R = 7.7$, while $R= 9.1$
and $6.0$ were obtained for the fainter and brighter NXB levels,
respectively.  The hard X-ray flux is also affected by NXB
uncertainties. 15--40 keV fluxes of $(2.3, 2.0, 2.5)\times10^{-11}$
erg s$^{-1}$ cm$^{-2}$ were obtained for the nominal, decreased, and increased
NXB, respectively.


\begin{figure}[t]
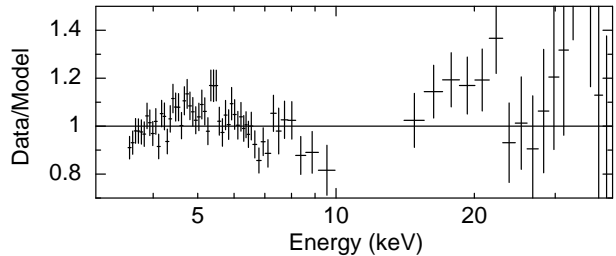

\begin{center}
\FigureFile(80mm,40mm){figure10.ps}
\caption{

Data/model ratio for a partial-covering model fit to the time-averaged
spectrum over the 3.5--4- keV band. The continuum is modeled using
a PEXRAV component with the reflection strength fixed at
$R$=1. For clarity, XIS 1 data are not plotted.
}
\end{center}
\end{figure}

\subsubsection{Complex Absorption Model Fit}


  Although the empirical model described above represents the observed
spectra, physical interpretation of the combination of the
very strong reflection continuum and moderate Fe-K emission line
($R \approx 6.8$ and $EW$ $\approx$ 140 eV for the average spectrum) is
not straight-forward, as discussed in Section 5.1.  We therefore
examined other continuum models for the constant hard spectral
component. First we assumed that the intrinsic continuum contains a
reflection component of $R = 1$, and that the apparent hard spectrum
is due to a partial covering absorber.  We fitted the averaged
spectrum with this partial covering model in the 3.5--40 keV band. The
PEXRAV component with fixed parameters $\Gamma = 2.04$ and $R = 1$ was
used as a continuum and partial covering absorption was applied.  The
other PEXRAV parameters, except for the normalization, were fixed
to the values shown in the previous subsection.  This model is
expressed as
\[
I(E) = e^{-\sigma (E) N_{\rm H, Gal}} [ \{ e^{-\sigma (E) N_{\rm H,1}} f  
\]
\[
~~~~~~~~~~~~ + (1-f) \} \{A C(E) + G_{\rm Fe}(E)\} ], 
\]
where $\sigma (E)$, $N_{\rm H, Gal}$, $N_{\rm H, 1}$, $f$, $A$,
$C(E)$, and $G_{Fe}(E)$ are the cross section of photoelectric absorption,
column density of the Galactic absorption, column density of the
intrinsic absorber, covering fraction, normalization factor of the
continuum, continuum model, and Gaussians to represent Fe-K emission
and absorption lines, respectively.  According to
this model, the very hard continuum is produced by a partially covered
continuum and the leaked emission dilutes the continuum around the
Fe-K line to reduce the apparent equivalent width.  Although the
equivalent width (140 eV) obtained in this fit and the assumed
strength of the reflection component ($R=1$) are consistent with each
other, this model resulted in systematic convex residuals in 3.5--7
keV (see Fig. 10).
The absorption column density and covering fraction obtained are
$\sim8.5\times10^{23}$ cm$^{-2}$ and $\sim0.67$, respectively. The
reduced $\chi ^2$ for this model is 1.66 for 124 degrees of freedom
(dof) for the averaged spectrum.  We added an additional absorbed
component to explain the convex residuals:
\[
I(E) = e^{-\sigma (E) N_{\rm H, Gal}} \{ e^{-\sigma (E) N_{\rm H,1}} f A C(E) 
\]
\[
~~~~~~ + (1-f) A C(E)  
\]
\[
~~~~~~ + e^{-\sigma (E) N_{\rm H,2}} B C(E) + G_{\rm Fe}(E) \},
\]
where 
$N_{\rm H,2}$ is the column density of the second absorber and $B$ is
the normalization of the additional continuum.  This model gave a good
fit to the averaged spectrum ($\chi^2_{\nu} = 1.10$ for 122 dof). 
The best-fit column densities  and covering fraction are 
$N_{\rm H,1} = 2.1\times10^{24}$ cm$^{-2}$, 
$N_{\rm H,2} = 5.2\times10^{22}$ cm$^{-2}$, 
and $f$ = 0.91, respectively. The ratio of the continuum normalizations
is $A/B \approx 2.2$. 
Although this model explains the spectral shape in 3.5--40 keV,
there are a few problems, as discussed in Section 5.1.


  If the small amplitude variability of the hard emission is due to
the significant contribution of a reflection component, we need a
self-consistent model explaining the strong reflection and the
strength of the Fe-K line.  One possible idea is that the $EW$ of the
Fe-K line is apparently small because of complex absorption. Hence we
assumed that a reflection continuum and an Fe-K line are partially
absorbed by cold material and that the $EW$ of the line is 1.0 keV
with respect to the reflection continuum. Note that an $EW$ of $\sim$
1 keV is expected for a reflection dominated spectrum (Guainazzi et
al.\ 1998, 2005; Bassani et al.\ 1999, Levenson et al.\ 2006).  In
this model, a PEXRAV component with {\tt rel\_refl} $ = -1$ was used
for the reflection component, and other parameters in PEXRAV were
treated as in Section 4.1. A power law with a fixed photon index of
2.04 was also added to these components. This model explains the
average spectrum very well, but failed to reproduce the shape of the
low flux state spectrum, namely the overall flat spectrum and the
convex shape in the 3.5--7 keV range. Such a spectral shape is not
obtained using this model even if a flatter photon index is allowed.
If the power-law component is also partially covered by absorber with
distinct parameters (covering fraction and column density), then the
spectral shape is explained.  This model is expressed using the
following equation:

\[
I(E) = e^{-\sigma (E) N_{\rm H, Gal}} [ \{ e^{-\sigma (E) N_{\rm H,1}} f_1  + (1-f_1) \} \{ A C(E) \} 
\]
\[ ~~~~~~~+  \{  e^{-\sigma (E) N_{\rm H,2}} f_2  + (1-f_2)  \}
\{ B R(E) + G_{\rm Fe}(E) \} ],
\]
where $C(E)$ is a power-law continuum, $f$ and $N_{\rm H}$ are the
covering fraction and column density of the partial absorber, and the
subscripts 1 and 2 denote the absorber for the power-law and
reflection components, respectively. $A$ and $B$ are the
normalizations of the power-law and reflection component,
respectively, and $G_{\rm Fe}(E)$ is Gaussians to
represent the Fe-K emission abd absorption lines. This model
(hereafter referred to as Model-B1) is explored more quantitatively
using the full energy band in the next subsection.

\subsection{Full-band Spectra}

If we extrapolate the best-fit model (Model-A1 or Model-B1) obtained
in the previous subsection, the data points are located below the
model in the 0.7--3.5 keV band and indicate the presence of a
complex absorber and an additional continuum component. The small separation
between the data and model at energies below 0.7 keV suggests that the
absorber is ionized and that there is an additional soft emission
component. The spectra also show absorption features due to ionized O, Ne, and
Mg in the 0.5--1.5 keV band, and due to Fe around 7 keV. 
Such features are indicative
of the presence of a multi-zone ionized absorber.  We therefore added two
ionized absorber components and a blackbody (BB) to model the soft
band spectra. We made a grid of ionized absorption usable in XSPEC by
using XSTAR in the HEASOFT6.4.1 package. This model assumed a
turbulent velocity of $\sigma = 200$ km s$^{-1}$, an ionizing
continuum with a photon index of 2.0, and solar abundances as in Grevesse
\& Sauval (1998). The blackbody component was introduced as an
empirical representation of the soft excess continuum. We fixed its
temperature at $kT$ = 0.12 keV, which is typical of blackbody temperatures
measured in Seyfert
1 galaxies (Pounds et al.\ 2004). Two baseline continuum models, the
models A1 and B1 in the previous section, were used. The photon index
was fixed at 2.04 unless otherwise noted.  Since several narrow
emission line features are present, we added Gaussian components to represent
them. The widths of the Gaussians were fixed at $\sigma$ = 10 eV
unless otherwise noted, while the width of the Fe-K line was left
free.


As noted above, 
an Fe-K$\beta$ emission line is not clearly seen in the spectra. If
an Fe-K$\beta$ line at 7.06 keV (source rest frame) was added to the
model with an intensity set to 0.13 times that of the Fe-K$\alpha$ line 
(Palmeri et al.\ 2003), significant line-like residuals remained around
7.0 keV, suggesting the presence of absorption around this
energy due to highly-ionized Fe. 
We therefore allowed the redshift parameter for the
more highly-ionized absorbing component to vary, and 
obtained a good fit if the absorption lines are allowed to be
blueshifted. After fixing 
the blueshift of the highly ionized absorber
at the best-fit values, this allowed us to then add a 
Fe-K$\beta$ line with a normalization set to
0.13 times that of the Fe-K$\alpha$ line.


We fitted the three
spectra (average, high flux, and low flux) with Models A1 and B1
combined with the ionized absorbers, the BB, and the Gaussians 
(henceforth referred to as Model-A2 and Model-B2, respectively).
Although Model-A2 is probably physically unreasonable, we examined
this model first because the number of model parameters is smaller
than the other complex models, and the parameters for discrete features are
reasonably well-constrained.  Model-A2 provided a good fit to the averaged
spectra.  The best-fit parameters are summarized in
Table 2. The possible identifications of the emission lines
are also shown in Table 2.
Fitting the low flux spectrum with this model resulted in
convex residuals in 2--6 keV, which suggests the presence of an
additional highly absorbed component. We therefore added an absorbed
power-law component with a fixed photon index of 2.04. This model yielded
a good fit, as summarized in Table 2. Alternatively, we examined a
broad relativistic Fe-K line, instead of the absorbed
component, as a possible explanation for the curved spectral shape in the 
2--6 keV band.  The LAOR line model in XSPEC (Laor 1991) was used for the broad
Fe-K line and we fixed the line energy, inner radius, outer
radius, emissivity index $q$ (where the emissivity $\propto r^{q}$), and
inclination at 6.4 keV (source rest frame), 1.235$R_{\rm g}$,
50$R_{\rm g}$, $-3$, and 30$^\circ$, respectively, where $R_{\rm g}$
is the gravitational radius, $GM_{\rm BH}/c^2$.
This model failed to explain the spectrum ($\chi^2$/dof = 668/427)
and systematic convex residuals remained
in the 3--6 keV band.  The best-fit equivalent
width of the broad and narrow lines were 120 and 140 eV, respectively,
which are too small given the strong reflection ($R \sim 7$) for this
model.


Next, we examined Model-B2, expressed in the following form:
\[
I(E) = e^{-\sigma (E) N_{\rm H, Gal}} {\rm IA}_1(\xi_1, N_{\rm H 1}) 
{\rm IA}_2(\xi_2, N_{\rm H 2}) 
\]
\[
~~~~~~
\times [ \{ e^{-\sigma (E) N_{\rm H,1}} f_1  + (1-f_1) \} \{ A C(E)  + G(E) \} 
\]
\[
~~~~~~ +  \{ 
e^{-\sigma (E) N_{\rm H,2}} f_2  + (1-f_2) 
\}
\{
B R(E)
+ G_{\rm Fe}(E)
\}
],
\]
where IA$_1$ and IA$_2$ are ionized absorbers, $G(E)$ is a combination
of Gaussians peaking at several different energies, and $G_{\rm
Fe}(E)$ is two Gaussians to represent the Fe-K$\alpha$ and $K\beta$
emission lines.  The parameters of the warm absorbers include the
ionization parameter $\xi \equiv L/nr^2$ (where $L$ is the isotropic,
ionizing luminosity in the 1--1000 ryd range, $n$ is the electron
number density, and $r$ is the distance from the central continuum
source to the absorbing gas), the column density, and the redshift
$z$.  Other components are the same as in Model-B1 in the previous
subsection.  The $EW$ of the Fe-K line with respect to the reflection
continuum is assumed to be 1.0 keV.  We fitted the three spectra with
this model (Model-B2) and obtained good fits to the spectra. The
best-fit parameters are shown in Table 3.


\begin{figure}[h]
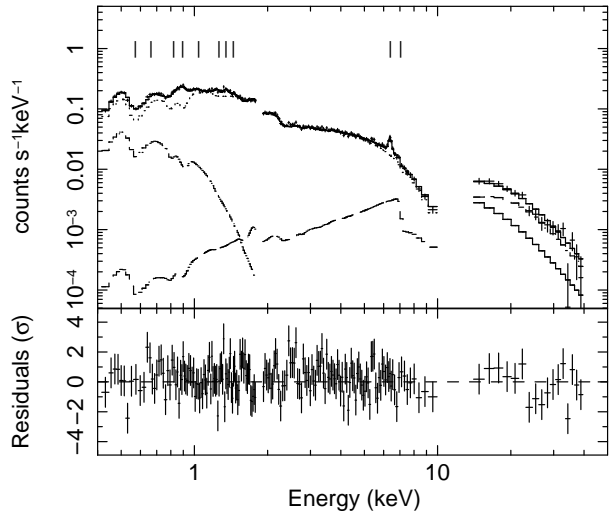

\begin{center}
\FigureFile(80mm,50mm){figure11.ps}
\caption{
The averaged spectrum fitted with the complex absorber model in the
0.4--40 keV band. Crosses denote the data and the solid histogram
denotes the best-fit model. Dotted, dot-dashed, and dashed lines denote the 
partially covered power-law component, the blackbody component, 
and partially covered reflection, respectively, with all 
components modified by ionized absorbers.
The best-fit energies of Gaussian components are marked.
The lower-panel shows the residuals.
}
\end{center}
\end{figure}

\begin{figure}[h]
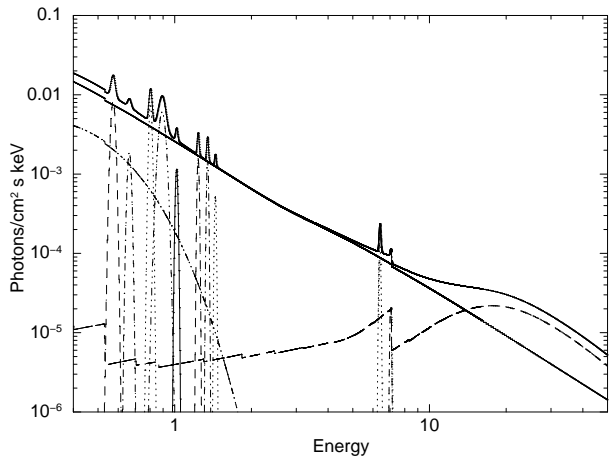

\begin{center}
\FigureFile(80mm,50mm){figure12.ps}
\caption{
The best-fit model components to the averaged spectrum shown in
Fig.\ 11. The ionized absorbers are removed. The normalization of the
models in the energy band of the HXD/PIN is set to unity to connect the
model lines in the low and high energy bands for the purpose of
presentation.
}
\end{center}
\end{figure}


\begin{figure}[h]
\begin{center}
\FigureFile(80mm,50mm){figure13.ps}
\caption{
The spectrum in the high flux state fitted with the complex 
absorber model in the 0.4--40 keV band. 
The lower-panel shows the residuals.
Symbols are same as in Fig.\ 11.
}
\end{center}
\end{figure}

\begin{figure}[h]
\begin{center}
\FigureFile(80mm,50mm){figure14.ps}
\caption{
The best-fit model components to the spectrum in the high flux state
 shown in
Fig.\ 13. The ionized absorbers are removed. The normalization of the
models in the energy band of the HXD/PIN is set to unity to connect the
model lines in the low and high energy bands for the purpose of
presentation.
}
\end{center}
\end{figure}

\begin{figure}[h]
\begin{center}
\FigureFile(80mm,50mm){figure15.ps}
\caption{
The spectrum in the low flux state fitted with the complex 
absorber model in the 0.4--40 keV band. 
The lower-panel shows the residuals.
Symbols are same as in Fig.\ 11.
}
\end{center}
\end{figure}

\begin{figure}[h]
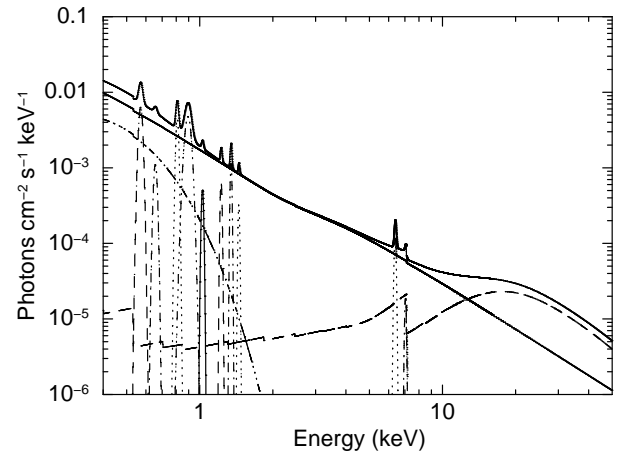

\begin{center}
\FigureFile(80mm,50mm){figure16.ps}
\caption{
The best-fit model components to the spectrum in the low flux state
 shown in
Fig.\ 15. The ionized absorbers are removed. The normalization of the
models in the energy band of the HXD/PIN is set to unity to connect the
model lines in the low and high energy bands for the purpose of
presentation.
}
\end{center}
\end{figure}

The best-fit spectrum and incident model spectrum for the averaged
spectrum are shown in Fig.\ 11 and Fig.\ 12, respectively. The
best-fit spectra and incident model spectra for the high and low
fluxes are displayed in Figs.\ 13, 14, 15, and 16, respectively.
X-ray fluxes in the 0.5--2, 2--10, and 15--40 keV bands for the
best-fit model are listed in Table 4.

In the spectral fits shown above, the power law photon index and the
temperature of the blackbody components were fixed at 2.04 and 0.12
keV, respectively. We allowed one of these two parameters to vary to
test whether these assumptions are appropriate. We obtained a photon
index of $2.042\pm0.008$, $2.041\pm0.012$, and
$2.038^{+0.007}_{-0.008}$ for the averaged, high flux, and low flux
spectra, respectively.  These results support that the photon index
remained almost constant and varied only in its normalization.  The
temperature of the blackbody component was $kT = 0.122\pm0.005$,
$0.121\pm0.011$, and $0.1176\pm0.0004$ keV, for the averaged, high
flux, and low flux spectra, respectively. Thus $kT$ also remained
nearly constant.

\section{Discussion}

\subsection{Two-component Behavior of Spectral Variability}

 NGC 4051 showed significant spectral variability during our {\it Suzaku}
observation. The spectrum hardens when the source becomes faint and
the amplitude of variability is smaller towards higher X-ray 
energies. This trend
of variability is similar to that observed in this object and many
other AGNs. Our time-resolved broad-band spectra showed a similar
trend at high energies ($>$2 keV), and can be used to distinguish among
various possibilities for the origin of the spectral variability.

\subsubsection{The Variable Power-law Component}

The spectral variability observed at energies above $\sim$3.5 keV is
well explained by the two-component model consisting of a variable
power-law component and a nearly constant hard component. We modeled the hard
component by a reflection continuum (Model-A1, A2) or a partially
covered power law plus a partially covered reflection continuum
(Model-B1, B2).  The photon indices of the power-law component for the
difference spectrum, and spectra in the high and low flux states agree
with each other ($\Gamma \approx 2.04$). This means that only the
normalization of the power-law component varied and that the photon
index remained almost constant.  Such a model of variability was
suggested by Ponti et al.\ (2006) based on {\it XMM-Newton} observations of NGC
4051.  Similar variability has been observed in MCG--6-30-15 (Miniutti et
al.\ 2007), MCG--5-23-16 (Reeves et al.\ 2007), etc.  This
behavior is also consistent with the linear shape of the flux-flux
plots in the hard energy bands (Fig.\ 4(b) and (c)).  Therefore, at least
in our observation, spectral pivoting (variation in the photon index) is
unlikely as the origin of the spectral variability.

\subsubsection{The Constant Hard Component}


  The spectral shape of the constant hard component can be fitted by
different models. We first tried to model this component as a
continuum consisting of reflection off of cold matter (models A-1 and A-2). 
This model resulted in very strong reflection ($R \approx$ 7 for the
averaged spectrum). A strong reflection with $R>1$ could be explained
in terms of the effect of light travel time.  If the reflector is
located far from the nuclear X-ray source, and if the the source
becomes faint compared to the historical average luminosity, $R$ can
be large. A problem in this model is that the $EW$ of the Fe-K line is not
compatible with the large value of $R$.  If the geometry of the reflector
is assumed to be a cold infinite slab, i.e., $R=1$, the $EW$ for the Fe-K
fluorescent line is expected to be 100--150 eV, unless the viewing
angle is almost edge-on (George \& Fabian 1991). The observed $EW$ should
be $R=7$ times these expected $EW$ values, while the observed value is 
$EW \approx$ 140 eV. Thus the interpretation that the constant component 
is purely from a reflected continuum is unlikely.
A component with a convex shape seen in the low flux state is an
additional piece of evidence that the shape of the constant 
component component is not a simple reflection continuum.


In order to solve the apparent inconsistency between the strength of
reflection $R$ and the Fe line $EW$, we examined a
partial covering model with the PEXRAV component having 
$R$ fixed at 1. This model failed to
explain the spectrum, and an additional component, an
absorbed power law, was required. This model, however, also has difficulties.
First, if the best-fit model is extrapolated to lower energies, a very
strong excess is seen below 3 keV, which is unusual for spectra of
Seyfert galaxies.  Secondly, if this idea is correct, X-rays at higher
energies must be less-absorbed intrinsic emission. Then it is required
that the X-ray source show intrinsic variability with small amplitude
and that the observed variability at lower energies is produced 
mostly by the behavior of the complex absorber. Such 
intrinsic variability with small amplitude 
is inconsistent with the trend that AGNs with
relatively smaller central black hole masses 
show more rapid and larger amplitude variability
(e.g., O'Neill et al.\ 2005).

We assumed that the small amplitude variability at higher energies is
due to a significant contribution from reflected emission from matter
distant from the nuclear X-ray source. In order to reconcile the
strength of the reflection component with the Fe-K $EW$, we introduced a
partial covering absorber to a combination of reflection and Fe
emission. The $EW$ of the
Fe line with respect to the pure reflection continuum is assumed to be
1 keV.  This model explains the observed spectra if the power-law
emission is also absorbed by a partial coverer, which has 
parameters distinct from those for the reflection+Fe 
component (Model-B1, B2).
The resulting $R$ values (the relative strength of the reflection
continuum with respect to the power-law component) are 4.3, 2.9, and
5.7 for the averaged, high flux, and low flux spectra, respectively. These
numbers are still very large compared to the value expected for a
reflector covering a solid angle of $2\pi$ viewed from the central
X-ray source ($R=1$). Such strong reflection could be due to an effect
of light travel time.  The observed flux in the 2--10 keV band
$8.1\times10^{-12}$ erg s$^{-1}$cm$^{-2}$ is factor of $\sim 4$ lower than the
flux averaged over 8 yrs, as obtained with {\it RXTE} PCA ($\sim
3\times10^{-11}$ erg s$^{-1}$ cm$^{-2}$) (M$^{\rm c}$Hardy et al.\ 2004). The
observed $R$ value is consistent with this decline of the nuclear
flux, if the reflector is located at several light years from the
nucleus. The origin of the reflector could be identified with the 
inner wall of a putative obscuring torus.  In this interpretation, the reflected
continuum is partially covered by a cold absorber. If the absorbing
matter in a torus is patchy, such a partial coverer might be
explained. The normalization and the covering fraction of the
reflection component remained constant. This fact is
consistent with the two-component picture consisting of a variable
power-law component and a constant hard component, where the latter is 
the partially covered reflection component.

The power-law component is also assumed to be partially covered in our
model. The power law emission is most likely emitted from the inner
region of the accretion disk, and the size of the emitting region could
be much smaller than the patchy clouds in a torus. The partial coverer
responsible for the power-law component should consist of clouds with
a size smaller than the emitting region.  A candidate for such clouds
is high density, low ionization clumps in a accretion disk wind, which
could be identified with ionized absorbers.  Although we assumed this
absorber is neutral in the spectral fits, a weakly ionized absorber can
also explain the spectrum.  A spectral component with a convex shape
represented by a power law partially absorbed by a column of gas
with $N_{\rm H}$ $\sim10^{23}$
cm$^{-2}$ is occasionally observed in Seyfert 1s when the source flux
becomes low (NGC 3227, George et al.\ 1998; Mrk 335, Grupe et al.\ 2007,
2008) and a similar process could be at work in NGC 4051.

\begin{figure}[h]
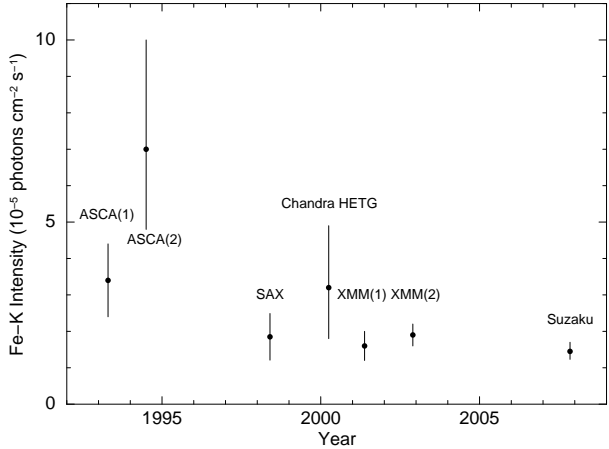

\begin{center}
\FigureFile(80mm,50mm){figure17.ps}
\caption{ History of Fe-K line intensity. Data are taken from Leighly
(1999), Lamer et al.\ (2003), Yaqoob \& Padmanabhan (2004), Pounds et
al.\ (2003), and this work for {\it ASCA}, {\it BeppoSAX}, {\it Chandra} HETG,
{\it XMM-Newton} PN, and {\it Suzaku}, respectively.  }
\end{center}
\end{figure}

\subsubsection{Long-Term Variability of the Hard X-ray Continuum and 
the Fe-K Emission Line}

If the Fe-K emission and reflection continuum are mainly emitted from
matter distant from the nucleus, their strengths are expected to be
constant over time scales of
years. Fig.\ 17 summarizes the observed Fe-K line intensities
obtained from past observations. Data from detectors with a moderate or good
spectral resolution are shown. The errors correspond to the interval
$\Delta \chi^2 =2.7$ except for the {\it BeppoSAX} data point by Lamer et al.\
(2003), given at a 1$\sigma$ level. The Fe-K line intensities are
nearly constant over 10 years and consistent with the idea that a
significant fraction of the Fe-K line is emitted from distant
matter. The second {\it ASCA} data point shows a larger error than others.  In
this fit, the line was broad ($\sigma = 0.5$ keV) (Leighly 1999).  A
possible interpretation is that a broad Fe-K component originating from
the inner part of accretion disk contributes to the flux and/or that
uncertainties in continuum modeling artificially make the line seem broader
and stronger.

If reflected emission makes a significant contribution to the hard
X-ray band as we have measured, hard X-ray fluxes are also expected to
remain nearly constant.  The history of hard X-ray fluxes is
summarized in Table 5.  We calculated hard X-ray fluxes in various
energy bands by using our averaged {\it Suzaku} spectrum and the best-fit
complex absorption model (Model-B2), and compared with published hard
X-ray measurements. The variations in hard X-ray fluxes among
{\it INTEGRAL}, {\it Swift}, and {\it Suzaku} observations are less than $\sim$50\%,
although assumptions on the spectral shape and cross calibration
between the instruments introduce large uncertainties.  For example,
Beckmann et al.\ (2006) measured a photon index of $\Gamma = 2.62$ in
the hard X-ray band with {\it INTEGRAL} ISGRI. This index is incompatible
with the {\it Suzaku} PIN data, which gives $\Gamma =
1.46^{+0.31}_{-0.22}$ if a simple power law is assumed. If $\Gamma
=2.62$ was assumed to fit the {\it Suzaku} PIN spectrum, reduced $\chi^2
=3.8$ (16 dof) and a 20--40 keV flux of $1.1\times10^{-11}$
erg s$^{-1}$cm$^{-2}$, which is about 60\% lower than the value for the
best-fit model, were obtained. Thus we conclude that hard X-ray fluxes
did not significantly change among {\it INTEGRAL}, {\it Swift}, and 
{\it Suzaku} measurements.

The hard X-ray flux obtained with {\it BeppoSAX} is $7.9\times10^{-12}$
erg s$^{-1}$ cm$^{-2}$ in the 15--40 keV band if the model given in Guainazzi et
al.\ (1998) is assumed.  This flux is about a factor of 3 lower
than that obtained with {\it Suzaku} ($2.4\times10^{-11}$ erg s$^{-1}$ cm$^{-2}$).
The {\it BeppoSAX} observation was done when the source was in a ultra-dim
state and the direct power-law component could have been switched off. If we
set the normalization of the power-law component to zero in our
{\it Suzaku} spectrum, the 15--40 keV flux becomes $1.8\times10^{-11}$
erg s$^{-1}$ cm$^{-2}$. Although this value is still about a factor of 2
larger than the {\it BeppoSAX} value, we consider the change between the two
observations to not be significant since the error bars at high energies
in the spectrum obtained with the PSD are relatively large.

The width of the Fe-K fluorescent line is expected to be narrow if the
reflector is located far from the nucleus.  The line width we obtained
is consistent with being narrow and agrees with the idea that the
contribution from distant matter is significant.  The line width
measured with {\it Chandra} HETG is $6330^{+131140}_{-4550}$ km
s$^{-1}$ at the 90\% confidence level for three parameters of interest
(Yaqoob \& Padmanabhan 2004).  
The lower bound of the error range is larger than the
width of the optical or UV broad lines
($1110\pm190$ km s$^{-1}$ for variable broad H$\beta$ and
$1040\pm250$ km s$^{-1}$ for CIV$\lambda1549$; Peterson et al.\
2000, Collinge et al.\ 2001).
The line profile shows a prominent narrow core and a broad residuals
in 5--6.4 keV (Fig.\ 4 in Yaqoob \& Padmanabhan 2004).  The
combination of these two components presumably resulted in the very
large error range, and a narrower core width would be allowed if a
better quality high-resolution spectrum is used to distinguish
multiple components.



\subsection{The Effect of Absorption}

The difference spectrum above 3.5 keV is fitted with a pure power-law component,
while the low energy part of the data shows a large deficit when the
power law is extrapolated from higher energies.  The hard X-ray spectrum
in the low flux state shows a component with a convex shape, 
represented by a partial covering model (Model-B2). These facts
suggest that variability of the absorbing components 
is another important factor in the spectral variability.
In Model-B2, the power-law continuum is covered by gas with a column
density of $N_{\rm H}$ $\sim 1\times10^{23}$ cm$^{-2}$, and its covering factor
varies with the flux.  This change explains most of the spectral
variability below 3.5 keV.  In fact, if the normalization of the
power-law component is increased and the covering fraction is
decreased in the best-fit model for the low flux, the broad band
continuum shape in the high flux state is mostly recovered. Weak
positive residuals are seen around 0.65--0.7 keV and 1.2--1.8 keV,
which may be related to small changes of ionized absorbers.  Most of
the the Gaussian emission lines did not show significant variability,
and the equivalent widths were larger in the low flux state. These
results and the RMS spectra showing smaller amplitudes at energies of
several emission lines imply that a significant fraction of the
emission line fluxes originate in an extended region, likely
photo-ionized by the emission from the central source. The low flux
state observed in 2002 with {\it XMM-Newton} also showed emission lines (Pounds
et al.\ 2003, Uttley et al.\ 2004, Ponti et al.\ 2006), which are also
compatible with the extended plasma origin of the lines.

Fig. 18 shows that the change of the power-law normalization and the
covering fraction can explain the most of the spectral variability
observed between the high and low flux states. In the following
procedure, only continuum components are considered.  We first changed
the normalization of the power-law component in the best-fit model in
the high flux state, setting it to the best-fit normalization for the
low flux state. This model was then divided by the best-fit model for
the high flux state. This ratio is shown as a dashed line in Fig.\
18. The ratio between the best-fit models in the low and high flux
states are shown as a solid line in Fig.\ 18.  The comparison between
these two lines shows that the spectral variability above 5 keV is
mostly explained by changing only the power-law normalization. Next,
we set both the normalization and the covering factor for the
power-law component at the best-fit values in the low flux state, and
divided this model by the best-fit model for the high flux state. This
ratio is shown as a dot-dashed line in Fig.\ 18, and it almost
reproduces the spectral shape in the low flux state.
Thus we conclude that the main causes of the spectral variability are
variations in the normalization of the power-law component with a
constant photon index, and variations in the covering fraction of the
absorber, which is assumed to be neutral in the fit and could be
weakly ionized. Additional contributors to the observed spectral
variability include small-amplitude changes of the warm absorbers and
the normalization of the BB component.

\begin{figure}[h]
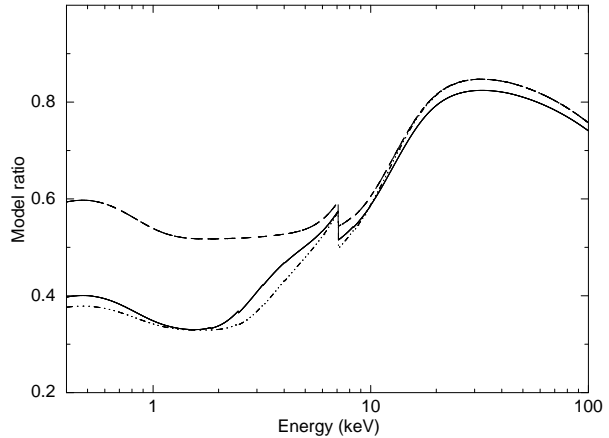

\begin{center}
\FigureFile(80mm,50mm){figure18.ps}
\caption{ 
Contribution of the change of the power-law component to the spectral
variability. Solid line: Ratio between the best-fit models in the low
and high flux states.  Dashed-line: The best-fit model in the high
flux state with the best-fit power-law normalization in the low flux
state, divided by the best-fit model in the high flux state.
Dot-dot-dot-dashed line: The best-fit model in the high flux state
with the best-fit power-law normalization and covering fraction in the
low flux state, divided by the best-fit model in the high flux state.
}
\end{center}
\end{figure}

\section{Summary and Conclusions}

  We analyzed data from a {\it Suzaku} observation of NGC 4051
obtained in 2005 Nov.\ and studied broadband spectral variability.
The averaged 2--10 keV flux was $8.1\times10^{-12}$ erg s$^{-1}$
cm$^{-2}$, which is several times lower than the historical average
($\sim 3\times10^{-11}$ erg s$^{-1}$ cm$^{-2}$) and larger than the
lowest flux observed in 1998 ($1.3\times10^{-12}$ erg s$^{-1}$
cm$^{-2}$).  NGC 4051 showed large amplitude variability during our
observation.  Several flare-like increases in the X-ray flux were
observed. The spectra became flatter when the source flux was low. We
examined the averaged spectrum and spectra in high and low flux time
intervals and identified variable components.

  The difference spectrum in the 3.5--40 keV band is well fitted by a
power-law component with a photon index of 2.04. This photon index is
in agreement with those measured for the high/low flux spectra, and
suggests that variations in the normalization of a power law with a
constant photon index combined with a nearly constant hard component
is the primary cause of the spectral variability above 3.5 keV.

  We fitted the full-band spectra with a model consisting of a power
law, reflection continuum from cold matter, a blackbody, two zones of
ionized absorbing material, and several Gaussian emission lines. The
hard constant component is dominated by a reflection continuum
originating in cold matter (Model-A2), the amount of reflection was $R
\sim 7$ for the averaged spectrum, where $R = 1$ corresponds to the
situation that the infinite slab subtends a solid angle of $2\pi$ as
viewed from the illuminating source, and we assumed an inclination
angle of 45$^\circ$. The observed $EW$ of the Fe-K line (140 eV),
however, is too small for the strong reflection. The hard constant
component therefore cannot be explained only by a reflection continuum
component.  We adopted a combination of partially covered reflection
and partially covered power-law emission as a model for the hard
component, and successfully explained the amount of reflection, the Fe
line $EW$, and the variability of the overall spectral shape. In this
model, an additional cause of spectral variability is the change of
the covering factor for the power-law component. The large covering
fraction of the power-law component is justified by the spectral shape
in the low flux state, which shows a convex shape reminiscent of a
spectrum absorbed by $\sim 1\times10^{23}$ cm$^{-2}$. The large amount
of reflection is presumably due to a light travel time effect, since
the flux in our observation is a factor of $\sim 4$ lower than the
historically averaged flux.  The parameters for the ionized absorbers,
blackbody, and Gaussian emission lines were nearly constant, and their
contribution to spectral variability is very limited.  The constant
emission line intensities are consistent with an origin in extended,
ionized gas.

  In summary, we identified the causes of spectral variability as
follows.  At high energies ($>$3.5 keV) the primary cause of the
spectral variability is variations in the normalization of the 
power-law component with a constant photon index ($\Gamma = 2.04$) overlaid
on a nearly constant hard component.  The constant hard component is
interpreted as a partially covered continuum.  At lower energies, 
variations in the covering fraction of the gas absorbing 
the power-law continuum is an
additional cause of the spectral variability.  Variations in 
the warm absorber parameters and the normalization of the 
blackbody component do not significantly contribute
to the spectral variability.


\bigskip

We thank an anonymous referee for useful comments that improved the
manuscript.
We are grateful to all the members of the {\it Suzaku} team.  The authors
thank T. Kallman for writing and supporting the XSTAR software. This
work is partially supported by Grants-in-Aid for Scientific Research
(20740109).

{}

\onecolumn

\small

\begin{table}
\tiny
\caption{Spcetral Parameters for Emprical Two Component Model fitted in 3.5--40 keV$^*$.}
\footnotesize
\begin{center}
\begin{tabular}{cccc}
\hline \hline
Flux State   & Average  & High Flux & Low Flux \\
\hline
Power law    & \\
Photon index & 2.04(f)  & 2.04(f) & 2.04(f)\\
Normalization$^\dagger$ & $2.40 (2.35-2.50)\times10^{-3}$ & $4.73 (4.51-4.94)\times10^{-3}$ & $1.91 (1.84-1.98)\times10^{-3}$ \\  
Reflection$^\ddagger$   & \\
Normalization$^\S$ & $18.0 (17.0-19.2)\times10^{-2}$      & $18.4 (15.4-21.4)\times10^{-2}$     & $18.2 (17.2-19.1)\times10^{-2}$ \\
$R^\|$          & 7.7      & 3.88    & 9.68\\
Gaussians$^\#$ & \\
Energy (keV) & $6.397 (6.382-6.412)$ & $6.370 (6.308-6.426)$ & $6.389 (6.375--6.403)$ \\
Width (keV)  & 0.044 ($0-0.075$)& 0.077 ($0-0.15$)   & 0.029 ($0-0.059$)\\
Normalization$^{**}$& $1.45 (1.22-1.70)$  & $1.92 (1.11-2.74)$ & $0.126 (0.108-0.144)$\\
EW (eV)      & $140 (119-164)$ & $122 (71-174)$  & $135 (115-155)$\\
Energy (keV) & $6.82 (6.72-6.88)$ & ...     & $6.80 (6.76-6.84)$\\
Width (keV)  & 0.01(f)  & ...     & 0.01(f)\\
Normalization& $-0.38 (-0.54 - -0.21)$ & ...     & $-0.40 (-0.54 - -0.27 )$\\
EW (eV)      & $-40 (-57 - -22)$    & ...     & $-46 (-65 - -33)$\\
Energy (keV) & ...      & ...     & $7.10 (7.04 - 7.15)$\\
Width (keV)  & ...      & ...     & 0.01(f)\\
Normalization& ...      & ...     & $-0.27 (-0.40 - -0.14)$\\
EW (eV)      & ...      & ...     & $-40 (-59 - -21)$\\
\hline
$\chi^2_{\nu}$/dof & 123.6/122  & 66.2/72 & 140.5/116\\  
\hline
\multicolumn{4}{@{}l@{}}{\hbox to 0pt{\parbox{140mm}{\footnotesize
Notes.
\par\noindent
* (f) denotes fixed parameter. Numbers in parentheses are error ranges
for $\Delta \chi^2$ = 2.7.
\par\noindent
$\dagger$ Normalization of power law in units of photons
keV$^{-1}$ cm$^{-2}$ s$^{-1}$ at 1 keV in the 
source rest frame. 
\par\noindent
$\ddagger$ Inclination angle of 45$^\circ$ was assumed.
\par\noindent
$\S$  Normalization of reflection continuum in units
of photons keV$^{-1}$ cm$^{-2}$ s$^{-1}$ at 1 keV 
of the incident power law only in the
observed frame.
\par\noindent
$\|$ Ratio of power law normalization to normaliztion
of reflection.
\par\noindent
$\#$ Gaussian center energies are in the source rest frame.
\par\noindent
** $10^{-5}$ photons cm$^{-2}$ s$^{-1}$ in the line.
}\hss}}
\end{tabular}
\end{center}
\end{table}


\begin{table}
\begin{center}
\caption{Spcetral Parameters for Emprical Two Component Model$^*$.}
\footnotesize
\begin{tabular}{ccccl}
\hline \hline
Flux         & Average  & High Flux & Low Flux & Line ID \\
\hline
Power law + Reflection$^\dagger$        & \\
Photon index & 2.04(f)  & 2.04(f) & 2.04(f)\\
Normalization$^\ddagger$& 2.66 (2.64--2.68)$\times10^{-3}$ & 5.08 (5.02--5.13)$\times10^{-3}$  & 1.72 (1.70--1.73)$\times10^{-4}$\\
$R$          & 6.9 (6.6--7.1)     & 3.7 (3.2--4.0) & 8.7 (8.3--9.0) \\
\hline
Blackbody    & \\
$kT$ (keV)   & 0.12(f)  & 0.12(f)   & 0.12(f) \\
Normalization$^\S$ & 1.71 (1.59--1.83)$\times10^{-5}$ & 3.50 (3.15--3.92)$\times10^{-5}$ & 2.09 (2.01--2.20)$\times10^{-5}$\\
\hline
Power law& \\
Photon index & ...     & ...                     & 2.04(f)\\
Normalization$^\|$ & ... & ...                     & 7.1 (6.7--7.5)$\times10^{-4}$ \\
$N_{\rm H}$ ($10^{22}$ cm$^{-2}$)& ... & ...       & $53 (47-59)$\\
\hline
Ionized absorber (1)$^\#$ &\\
$\log \xi$   & 1.84 (1.83--1.85) & 1.76 (1.74--1.78)  & 1.84 (1.82--1.85)\\
$N_{\rm H}$ ($10^{22}$ cm$^{-2}$) & 1.42 (1.23--1.59)   & 1.60 (1.54-1.68) & 1.94 (1.90--2.00)\\
Ionized absorber (2)$^{**}$ &\\
$\log \xi$   & 2.64 (2.63--2.67) & 2.89 (2.54--2.94) & 2.98 (2.93--3.00)\\
$N_{\rm H}$ ($10^{22}$ cm$^{-2}$) & 1.43 (1.26--1.63)  & 1.5 (1.0--2.3) & 2.7 (1.8--3.4)\\
\hline
Gaussians$^{\dagger \dagger,\ddagger \ddagger}$ \\
Energy (keV) & 0.582 (0.577--0.587)  & 0.577 (0.563-0.589) & 0.567 (0.561--0.573)   & OVII K$\alpha$\\
Normalization$^{\S \S}$ & 22.7 (19.7--26.5) & 23.4 (13.7--32.6) & 16.5 (13.4--20.5)\\
EW (eV)      & 22      & 12      & 20\\
Energy (keV) & 0.675 (0.664--0.687) & 0.659 (0.650--0.667) & 0.653 (0.639--0.666)   & OVIII Ly$\alpha$\\
Normalization& 7.0 (5.1--8.9) & 23 (18--30) & 3.4 (1.7--5.0) \\
EW (eV)      & 9      & 16  & 6\\
Energy (keV) & 0.831 (0.824--0.837) & 0.827 (0.821--0.833) & 0.803 (0.794--0.813)   & Fe XVII L?\\
Normalization& 12 (10--14)     & 20 (16--24) & 5.3 (3.6--7.1)\\
EW (eV)      & 23     & 24        & 14 \\
Energy (keV) & ...  & ...                                    & 0.884 (0.880--0.887) & Fe XVIII L? \\
Normalization& ...  & ...                                    & 24 (22--25)\\
EW (eV)      & ...  & ...                                    & 66 \\
Energy (keV) & 0.916 (0.914--0.918)  & 0.898 (0.890--0.910) & 0.912 (0.905--0.920)  & Ne IX K$\alpha$\\
Normalization& 48 (46--51) & 16 (12--19) & 8.2 (6.6--9.8)\\
EW (eV)      & 127      & 22   & 20 \\
Energy (keV) & 1.037 (1.030--1.042)  & ... & 1.026 (1.006--1.044)  & Ne X Ly$\alpha$\\
Normalization& 5.1 (4.4--6.0) & ...     & 0.93 (0.33--1.5)\\
EW (eV)      & 18                 & ...     & 5 \\
Energy (keV) & ...    & ...                                  & 1.26 (1.24--1.27)  & Fe XXIII L?\\
Normalization& ...    & ...                                  & 1.5 (1.0--1.9)\\
EW (eV)      & ...    & ...                                  & 12\\
Energy (keV) & 1.38 (1.37--1.39)  & 1.35 (1.30--1.38) & 1.35 (1.34--1.36)   & MgXI K$\alpha$\\
Normalization& 2.5 (2.1--3.0)& 1.9 (0.6--3.3)$\times10^{-5}$ & 2.5 (2.1--2.8)\\
EW (eV)      & 17      & 6    & 25\\
Energy (keV) & 1.470 (1.445--1.490)   & 1.443 (1.416--1.474) & ...  & Mg XII Ly$\alpha$\\
Normalization& 0.79 (0.38--1.18) & 2.4 (1.3--3.6) & ...\\
EW (eV)      & 6     & 10   & \\
Energy (keV) & 6.534 (6.517--6.550) & 6.524 (6.469--6.597) & 6.393 (6.379--6.406) & Fe K$\alpha$ \\
Width (keV)  & 0.035 (0--0.059)   & 0.080 (0.022--0.156) & 0.034 (0.0--0.056)\\
Normalization& 1.51 (1.30--1.69) & 1.98 (1.29--2.67) & 1.41 (1.24--1.58)\\
EW (eV)      & 140     & 119      & 179\\
\hline
$\chi^2_{\nu}$ (dof) & 614.0/486  & 342.8/309 & 570.1/421\\
\hline
\multicolumn{5}{@{}l@{}}{\hbox to 0pt{\parbox{140mm}{\footnotesize
Notes.
\par\noindent
* (f) denotes fixed parameter. 
Numbers in parentheses are error ranges
for $\Delta \chi^2$ = 2.7.
\par\noindent
$\dagger$ Inclination angle of 45$^\circ$ was assumed.
\par\noindent
$\ddagger$ Normalization of reflection continuum in units
of photons
keV$^{-1}$ cm$^{-2}$ s$^{-1}$ at 1 keV 
of the incident power law only in the
observed frame.
\par\noindent
$\S$ 
Normalization of Blackbody defined as
$L_{45}/D_{\rm 10 Mpc}$, where $L_{45}$ is the
source luminosity in units of $10^{45}$  erg s$^{-1}$
and $D_{\rm 10 Mpc}$ is the distance to the
source in units of 10 Mpc.
\par\noindent
$\|$ Normalization of power law in units of photons
keV$^{-1}$ cm$^{-2}$ s$^{-1}$ at 1 keV in the 
source rest frame. 
\par\noindent
$\#$ Redshift parameter for the warm absorber model with a low ionization
parameter is fixed at the redshift of NGC 4051 ($z$ = 0.002336).
\par\noindent
** Redshift parameter for the warm absorber model with a high ionization
parameter is fixed at the value at the $\chi^2$ minimum ( $z = -0.0575,
-0.059$, and $-0.0551$ for average, high flux, and low flux spectra,
respectively.
\par\noindent
$\dagger \dagger$ Width of the Gaussian at 6.4 keV was left free.
Widths of other Gaussians were fixed at 10 eV.
\par\noindent
$\ddagger \ddagger$ Fe K$\beta$ line at 7.06 keV (source rest frame) with a
normalization of 0.13 times that of Fe K$\alpha$ line is
also added. The width of Fe K$\alpha$ and K$\beta$ lines
are assumed to be common.
\par\noindent
$\S\S$ $10^{-5}$ photons cm$^{-2}$ s$^{-1}$ in the line.
}\hss}}
\end{tabular}
\end{center}
\end{table}



\begin{table*}
\begin{center}
\caption{Spcetral Parameters for Complex Absorber Model$^*$.}
\footnotesize
\begin{tabular}{ccccl}
\hline \hline
Flux state   & Average  & High Flux & Low Flux  & Line ID\\
\hline
Pratially Covered Power law    & \\
Photon index & 2.04(f)  & 2.04(f) & 2.04(f)\\
Normalization$^\dagger$& 4.16 (4.13--4.19)$\times10^{-3}$ & 6.43 (6.35--6.50)$\times10^{-3}$ & 3.32 (3.29--3.35)$\times10^{-3}$ \\
$N_{\rm H}$ (10$^{22}$ cm$^{-2}$)& 11.9 (10.9--13.0) & 12.1 (8.6--18.7) & 7.7 (7.1--8.2)\\
Covering fraction & 0.360 (0.349--0.370) & 0.156 (0.122-0.187) & 0.464 (0.455--0.473)\\
\hline
Partially Covered Reflection$^\ddagger$  &   & \\
$R$          & -1(f) & -1(f) & -1(f) \\
Normalization$^\S$ & 1.78 (1.68--1.89)$\times10^{-2}$ & 1.94 (1.65--2.22)$\times10^{-2}$ & 1.89 (1.79--2.00)$\times10^{-2}$\\
$N_{\rm H}$ ($10^{22}$ cm$^{-2}$)& 100(f) & 100(f) & 100(f)\\
Covering fraction & 0.806 (0.775--0.834) & 0.786 (0.708-0.880) & 0.802 (0.776--0.825)\\
\hline
Blackbody    & \\
$kT$ (keV)   & 0.12(f)  & 0.12(f) & 0.12(f) \\
Normalization$^\|$ & 3.26 (3.10--3.49)$\times10^{-5}$ & $3.53 (3.06-4.00)\times10^{-5}$ & 4.10 (3.88--4.33)$\times10^{-5}$ \\
\hline
Ionized absorber (1)$^\#$ &\\
$\log \xi$   & 1.79 (1.78--1.80) & 1.79 (1.77-1.81) & 1.82 (1.800--1.83)\\
$N_{\rm H}$ ($10^{22}$ cm$^{-2}$) & 2.28 (2.23--2.31) & 1.90 (1.83--1.97) & 2.06 (2.01--2.10)\\
Ionized absorber (2)$^{**}$ &\\
$\log \xi$   & 2.96 (2.94--3.00) & 2.71 (2.68-2.75) & 2.97 (2.91--3.02)\\
$N_{\rm H}$ ($10^{22}$ cm$^{-2}$) & 2.74 (2.04--3.5) & 1.30 (1.02--1.80) & 1.27 (0.87--1.75)\\
\hline
Gaussians$^{\dagger\dagger, \ddagger\ddagger}$ \\
Energy (keV) & 0.572 (0.567--0.5767) & 0.574 (0.560--0.587) & 0.570 (0.5642--0.576)  & OVII K$\alpha$\\
Normalization$^{\S\S}$ & 36.3 (31.2--42.1) & 33.2 (21.5--45.3) & 33.3 (27.6--39.4)\\
EW (eV)      & 22       & 14 & 22\\
Energy (keV) & 0.664 (0.649--0.679) & 0.660 (0.654--0.667) & 0.658 (0.639--0.675) & OVIII Ly$\alpha$\\
Normalization& 8.1 (5.5--10.8)  & 2.6 (1.9--3.2) & 5.6 (2.6--8.6)\\
EW (eV)      & 7        & 15 & 5\\
Energy (keV) & 0.822 (0.818--0.827) & 0.841 (0.8346--0.8463) & 0.827 (0.821--0.832)  & Fe XVII L?\\
Normalization& 31 (27--34) & 26 (20--31) & 22 (19--25)\\
EW (eV)      & 38        & 23 & 32\\
Energy (keV) & 0.895 (0.893--0.897) & 0.899 (0.892-0.906) & 0.896 (0.894--0.899) & Ne IX K$\alpha$\\
Normalization& 63 (60--66) & 42 (35--47) & 58 (55--61)\\
EW (eV)      & 106     & 48 & 117 \\
Energy (keV) & 1.041 (1.027--1.056) & 1.032 (0.995--1.050) & 1.050 (1.024--1.066) & Ne X LyXS$\alpha$\\
Normalization& 4.7 (3.3--5.8) & 5.3 (2.5--7.9) & 2.4 (1.4--3.7)\\
EW (eV)      & 11       & 8 & 7\\
Energy (keV) & 1.265 (1.259--1.271) & 1.279 (1.253--1.307) & 1.253 (1.236--1.265) & Fe XXIII L?\\
Normalization& 6.3 (5.5--7.3) & 3.1 (1.4--5.0) & 3.0 (2.3--4.0)\\
EW (eV)      & 22        & 8 & 13\\
Energy (keV) & 1.349 (1.342--1.353) & 1.351 (1.325--1.363) & 1.349 (1.343--1.355)  & Mg XI K$\alpha$\\
Normalization& 6.5 (5.5--6.7) & 3.9 (2.1--5.7) & 5.4 (4.8--6.3)\\
EW (eV)      & 26       & 11 & 29\\
Energy (keV) & 1.447 (1.434--1.463) & 1.449 (1.426-1.477)& 1.449 (1.430--1.467) & Mg XII Ly$\alpha$\\
Normalization& 2.1 (1.4--2.7)& 2.3(0.8-3.7)& 1.6 (1.0--2.2)\\
EW (eV)      & 10       & 7 & 10\\
Energy (keV) & 6.400 (6.385--6.414) & 6.380 (6.323--6.435) & 6.393 (6.380--6.408) & Fe K$\alpha$\\
Width (keV)  & 0.0469 (0.018--0.067) & 0.052 (0--0.129) & 0.056 (0.036--0.077)\\
Normalization & 4.9 & 5.2 & 5.1 \\
EW (keV)$^{\|\|}$ & 1.0(f)  & 1.0(f) & 1.0(f)\\
\hline
$\chi^2_{\nu}$ (dof) & 585.0/480  & 341.0/302 & 525.9/420\\
\hline
\multicolumn{5}{@{}l@{}}{\hbox to 0pt{\parbox{140mm}{\footnotesize
Notes.
\par\noindent
* (f) denotes fixed parameter.
Numbers in parentheses are error ranges
for $\Delta \chi^2$ = 2.7.
\par\noindent
$\dagger$ Normalization of power law in units of photons
keV$^{-1}$ cm$^{-2}$ s$^{-1}$ at 1 keV in the 
source rest frame. 
\par\noindent
$\dagger$ Inclination angle of 45$^\circ$ was assumed.
\par\noindent
$\ddagger$ Normalization of reflection continuum in units
of photons
keV$^{-1}$ cm$^{-2}$ s$^{-1}$ at 1 keV 
of the incident power law only in the
observed frame.
\par\noindent
$\S$
Normalization of Blackbody defined as
$L_{45}/D_{\rm 10 Mpc}$, where $L_{45}$ is the
source luminosity in units of $10^{45}$ erg s$^{-1}$
and $D_{\rm 10 Mpc}$ is the distance to the
source in units of 10 Mpc.
\par\noindent
$\|$ Redshift parameter for the warm absorber model with a low ionization
parameter is fixed at the redshift of NGC 4051 ($z$ = 0.002336).
\par\noindent
$\#$ Redshift parameter for the warm absorber model with a high ionization
parameter is fixed at the value at the $\chi^2$ minimum ( $z = -0.061,
-0.059$, and --0.049 for average, high flux, and low flux spectra,
respectively.
\par\noindent
** Gaussian center energies are in the source rest frame.
\par\noindent
$\dagger\dagger$ Width of the Gaussian at 0.89 keV fixed at
25 eV. Width of the Gaussian at 6.4 keV was left
free.
Widths of other lines were fixed at 10 eV.
\par\noindent
$\ddagger\ddagger$ Fe K$\beta$ line at 7.06 keV (source rest frame) with a
normalization of 0.13 times that of Fe K$\alpha$ line was
also added. The width of Fe K$\alpha$ and K$\beta$ lines
are assumed to be common.
\par\noindent
$\S\S$ $10\times10^{-5}$ photons cm$^{-2}$ s$^{-1}$ in the line.
\par\noindent
$\|\|$ Equivalent width was fixed at 1 keV with respect to the pure
reflection continuum before affected by absorption.
}\hss}}
\end{tabular}
\end{center}
\end{table*}


\begin{table*}
\begin{center}
\caption{Observed X-ray Fluxes for the Complex Absorber Model}
\footnotesize
\begin{tabular}{lccc}
\hline \hline
Flux state  &  Flux (0.5--2 keV)   & Flux (2--10 keV)  & Flux (15--40 keV) \\
            & ($10^{-11}$ erg s$^{-1}$ cm$^{-2}$)  & ($10^{-11}$ erg s$^{-1}$ cm$^{-2}$)  & ($10^{-11}$ erg s$^{-1}$ cm$^{-2}$) \\
\hline
Average     & 0.39  & 0.82 & 2.4 \\
High flux   & 0.78  & 1.4  & 2.9 \\
Low flux    & 0.30  & 0.68 & 2.3 \\
\hline
\end{tabular}
\end{center}
\end{table*}

\begin{table}
\begin{center}
\caption{Histroy of Hard X-ray Flux}
\footnotesize
\begin{tabular}{cccccc}
\hline \hline
Date   & Instrument  & Flux   & Flux (15-40 keV)$^*$ & Energy Band$^\dagger$ & Reference$^\ddagger$\\
       &             & ($10^{-11}$ erg s$^{-1}$cm$^{-2}$) & ($10^{-11}$ erg s$^{-1}$cm$^{-2}$) & (keV) \\
\hline
1998 May 9--11 & {\it BeppoSAX} MECS+PDS$^\S$ & 0.79 & 2.4 & 15--40 & 1\\
2002 Dec. -- 2006 Jun. & {\it INTEGRAL} ISGRI & 2.11 & 3.2 & 17--60 & 2\\ 
2002 Oct. -- 2004 Jan. & {\it INTEGRAL} ISGRI & 1.80  & 1.9 & 20--40 & 3\\
2002 Oct. -- 2004 Jan. & {\it INTEGRAL} ISGRI & 1.98  & 1.5 & 40--100& 3\\
2005 Dec. -- 2006 Sep. & {\it Swift} BAT  & 4.6 & 5.3  & 14--195 & 4\\
2005 Nov. 10--13       & {\it Suzaku} HXD PIN & ... & 2.4 & 15--40 & 5\\
\hline
\multicolumn{5}{@{}l@{}}{\hbox to 0pt{\parbox{140mm}{\footnotesize
Notes.
\par\noindent
* 
Fluxes in the 15--40 keV band calculated by assuming
the  best-fit
complex absorber model shown in Table 3.
\par\noindent
$\dagger$
Energy band used in the reference.
\par\noindent
$\ddagger$
References: 
(1) Guainazzi et al. 1998;
(2) Sazonov et al. 2007, Krivonos et al. 2007; 
(3) Beckmann et al. 2006;
(4) Tueller et al. 2008
(5) This work.
\par\noindent
$\S$
Model 6 in Guainazzi et al. (1998)
(pure reflection plus single broad Gaussian model) was assumed.
}\hss}}
\end{tabular}
\end{center}
\end{table}

\end{document}